\documentclass[twoside]{dis07}
\usepackage[latin1]{inputenc}
\usepackage[dvips]{graphicx,epsfig,color}
\usepackage{wrapfig} 

\newcommand{\bm}[1]{\mbox{\boldmath $#1$}}

\newcommand{\simordertwo}{\raisebox{-3pt}{$\, \stackrel{ 
      <}{\sim} \,$}}

\begin{document}
\title{Spin Physics: Session Summary}

\author{Dani\"el Boer$^1$, Delia Hasch$^2$ and Gerhard Mallot$^3$
%
%
\vspace{.3cm}\\
%
1- Vrije Universiteit Amsterdam - Department of Physics and Astronomy\\
De Boelelaan 1081, NL-1081 HV Amsterdam - The Netherlands
%
\vspace{.1cm}\\
2- INFN Laboratory Nazionali di Frascati \\
Via Enrico Fermi 40, I-00044 Frascati - Italy
\vspace{.1cm}\\
3- CERN, CH-1211 Geneva 23 - Switzerland\\
}

\maketitle

\begin{abstract}
We summarize the main results of the spin physics Working Group Session of DIS
2007, the 15th International Workshop on ``Deep-Inelastic Scattering and 
Related Subjects''. 
\end{abstract}

\section{Introduction}

Many spin physics experiments have been performed in recent years and many new
exciting results have been reported at DIS 2007 \cite{url}, which will be
highlighted in this summary. Also on the theory side many new results were
reported, especially regarding transverse spin effects which are most
challenging. Recent years have seen quite some unexpected developments
concerning so-called TMDs, transverse momentum dependent parton distributions,
and we can look forward to more such developments over the coming years.
Therefore, this summary is very much a snapshot of the current status.

This summary is split into three main parts. We start with longitudinal spin
physics, most notably, experimental results on gluon polarization. We proceed
with transverse spin, which is mainly focused on transverse spin asymmetries
and the possible explanation in terms of TMDs. The third and last part is
about exclusive processes and generalized parton distributions, which provide
more detailed information about the spatial distribution of partons inside
hadrons. This spatial distribution is often probed using spin asymmetries and
recent developments have started to point to a connection between GPDs and
TMDs. A very interesting development. 

\section{Longitudinal spin}

The discovery by the European Muon Collaboration \cite{Ashman:1989ig} that 
the first moment $\Gamma_1^p$ of the spin-dependent structure function $g_1^p$
of the proton   
\[
\Gamma_1^{p/n}(Q^2) = \int_0^1 {\rm d}x \; g_1^{p/n}(x,Q^2) = \frac{1}{36} 
\left(4 \Delta \Sigma \pm 3 \Delta q_3 + \Delta
    q_8 \right) \left(1 + \frac{\alpha_s}{\pi}\right) + {\cal O}(\alpha_s^2)
\]
is much smaller than expected implies that the total contribution of the
quark spins to the nucleon spin
$\Delta \Sigma \equiv \Delta u + \Delta d + \Delta s$ is small.
Here
$\Delta q = (q_+ - q_-) + (\bar q_+ - \bar q_-) $ is 
the difference of the number of quarks and antiquarks of flavor $q$ with
positive and negative helicity and $\Delta q_3 \equiv \Delta u - \Delta d$ 
and $\Delta q_8 \equiv \Delta u + \Delta d -2\Delta s$ are known
from $\beta$ decays. 

HERMES presented the final analysis of their $g_1^p$ and 
$g_1^d$ measurements 
\cite{Airapetian:2007mh} and COMPASS showed new, very precise deuteron 
data \cite{Alexakhin:2006vx} (Fig.~\ref{fig:cmp_g1d}).
Both collaborations evaluated $\Delta\Sigma$ from their deuteron data with $Q^2>1~{\rm GeV}^2$ 
yielding
$\Delta \Sigma = 0.330 \pm 0.025\, (\mbox{exp.}) \pm 0.028\, (\mbox{evol.})
\pm 0.011\, (\mbox{theo.})$ at   $Q^2 = 5~\rm{GeV}^2$ from HERMES and 
$\Delta \Sigma = 0.35 \pm 0.03\,(\mbox{stat.}) \pm 0.05\, (\mbox{syst.})$
at $Q^2 = 3~\rm{GeV}^2$ from COMPASS. The results are in excellent agreement.
The value for $\Delta\Sigma$ is somewhat larger than the original EMC result of 
$\Delta \Sigma = 0.12\pm0.17$, which was given at a larger scale 
$Q^2=10.7\ \rm{GeV}^2$. All results are consistent with each other upon taking 
evolution into account. Therefore, the conclusion that the quark spins 
contribute little to the nucleon spin remains valid. 

CLAS from JLAB showed a wealth of proton and deuteron $g_1$ data covering the range
$0.05<Q^2<5~{\rm GeV}^2$. For $Q^2>1~{\rm GeV}^2$ the range $0.15<x<0.58$ is 
covered \cite{Griffioen} (Fig.~\ref{fig:CLAS}). The spin structure in the resonance
region and the Burkhardt--Cottingham sum rule were explored by the Hall-C experiment 
E01-006 \cite{Slifer}.

Semi-inclusive DIS (SIDIS), in which in addition to the scattered lepton a hadron 
is observed, can be analyzed in terms of the valence quark helicity distributions
$\Delta q_v$. New COMPASS deuteron data obtained in leading order (LO) and using a 
fragmentation-function independent method \cite{Korzenev} are shown in Fig.~\ref{fig:dq_valence} 
together with previous data. They disfavour a flavor-symmetric quark sea with
$\Delta\overline{u}=\Delta\overline{d}=\Delta s=\Delta\overline{s}$.

\begin{figure}
\begin{minipage}[b]{0.47\hsize}
\includegraphics[width=\columnwidth]{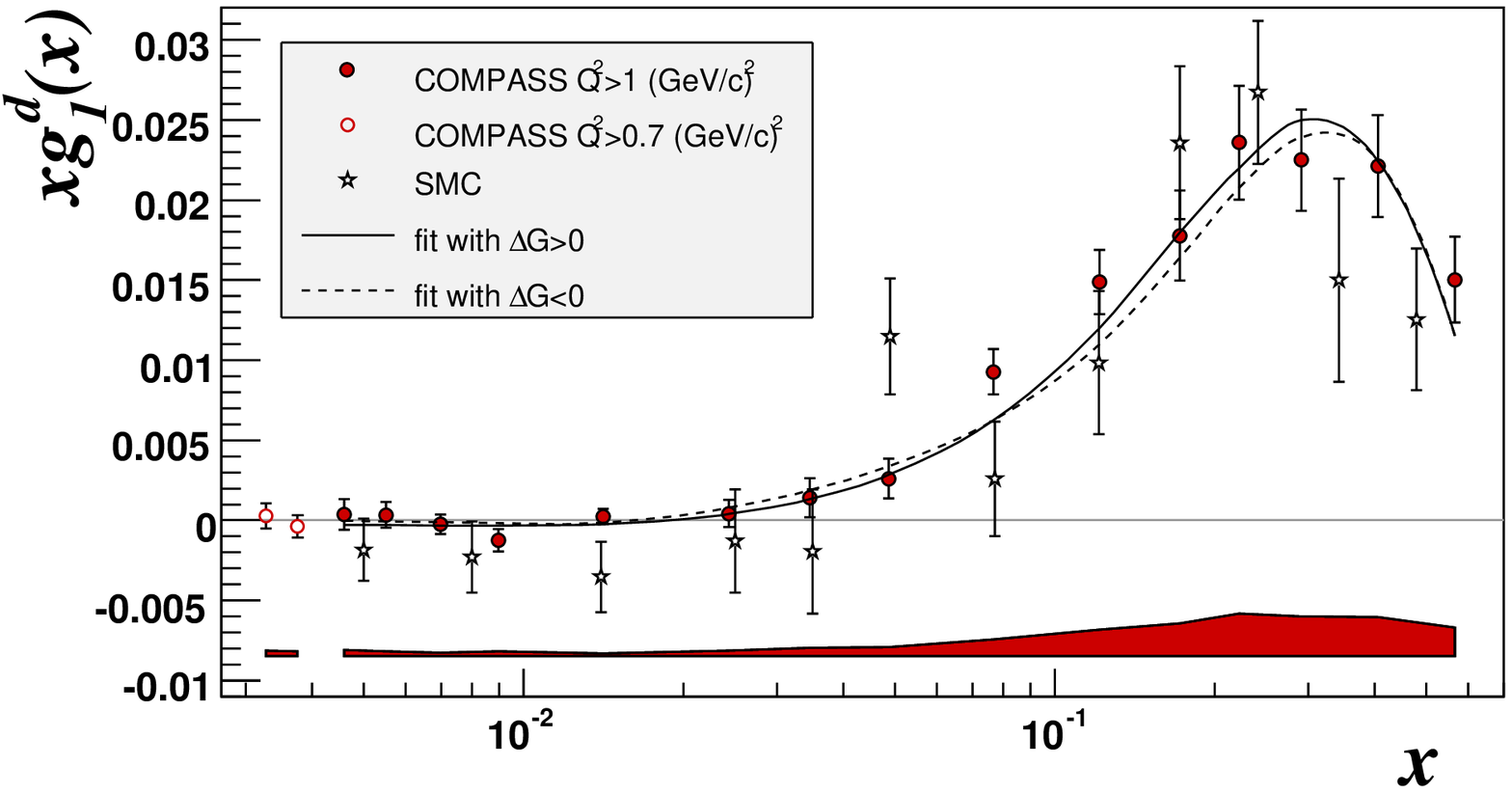}
\caption{\label{fig:cmp_g1d}The deuteron structure function $xg_1^d$ as function of $x$ 
from COMPASS \cite{Alexakhin:2006vx}. Also shown are QCD fits with positive and negative
$\Delta G$.}
\vspace{3mm}
\includegraphics[width=\columnwidth]{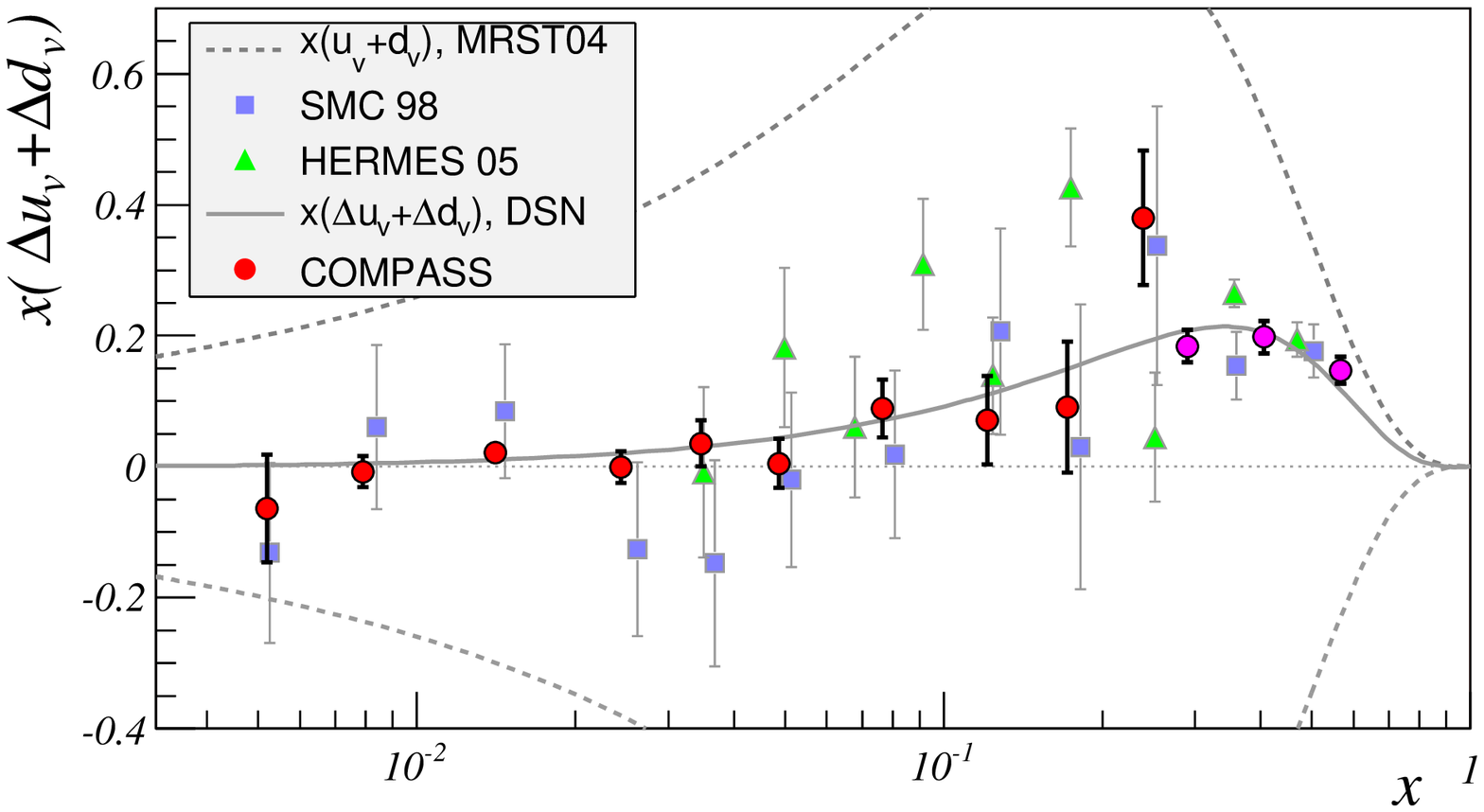}
\addtocounter{figure}{1}
\caption{\label{fig:dq_valence}The valence quark distribution $x(\Delta u_v + \Delta d_v)$ as function
of $x$ from SIDIS obtained in LO and evolved to $Q^2=10~\rm{GeV}^2$ \cite{Korzenev} using the PDFs 
of Ref.~\cite{deFlorian:2005mw}.}
\end{minipage}
\hfil
\begin{minipage}[b]{0.47\hsize}
\begin{center}
\includegraphics[width=0.8\columnwidth]{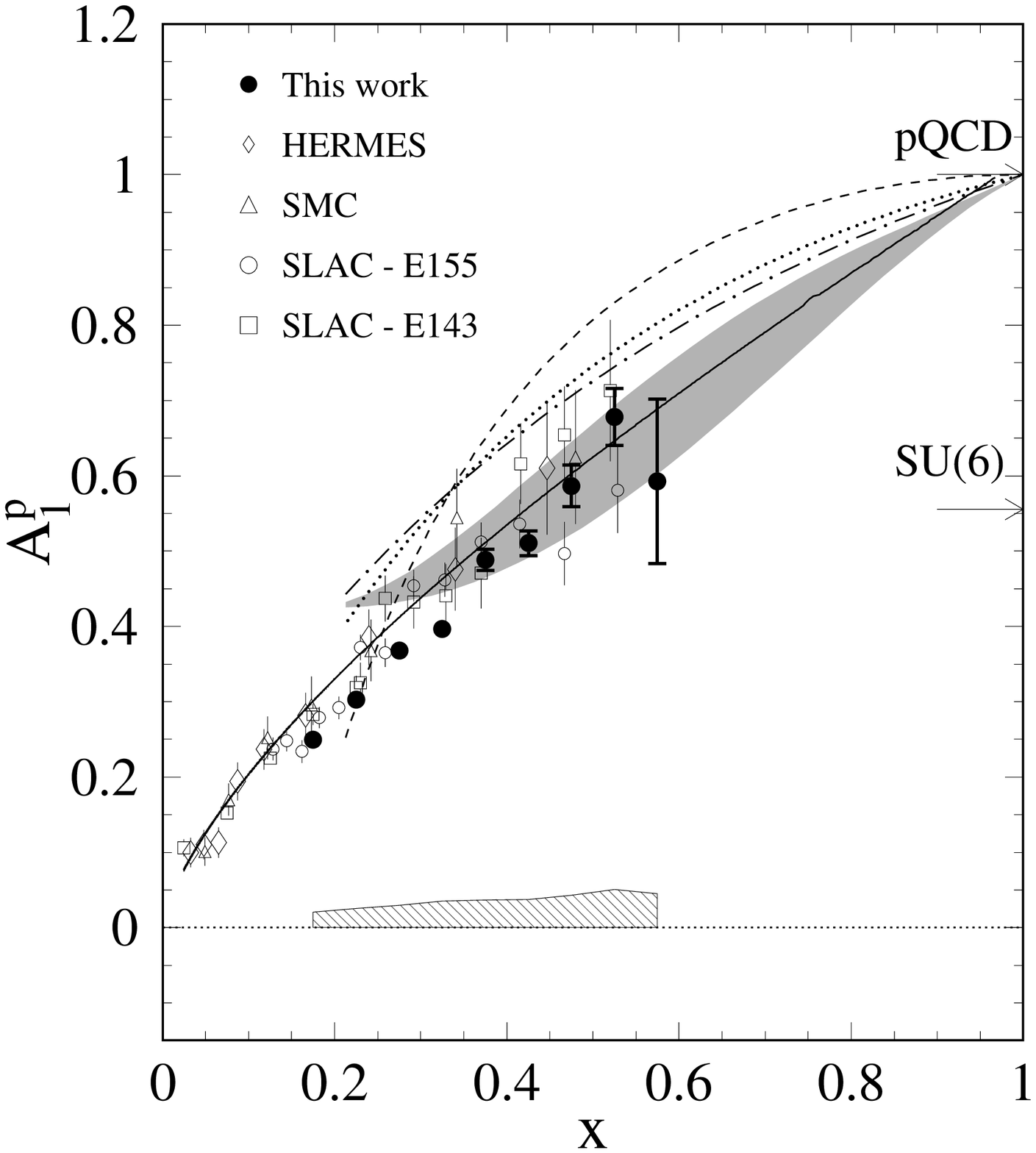}
\includegraphics[width=0.8\columnwidth]{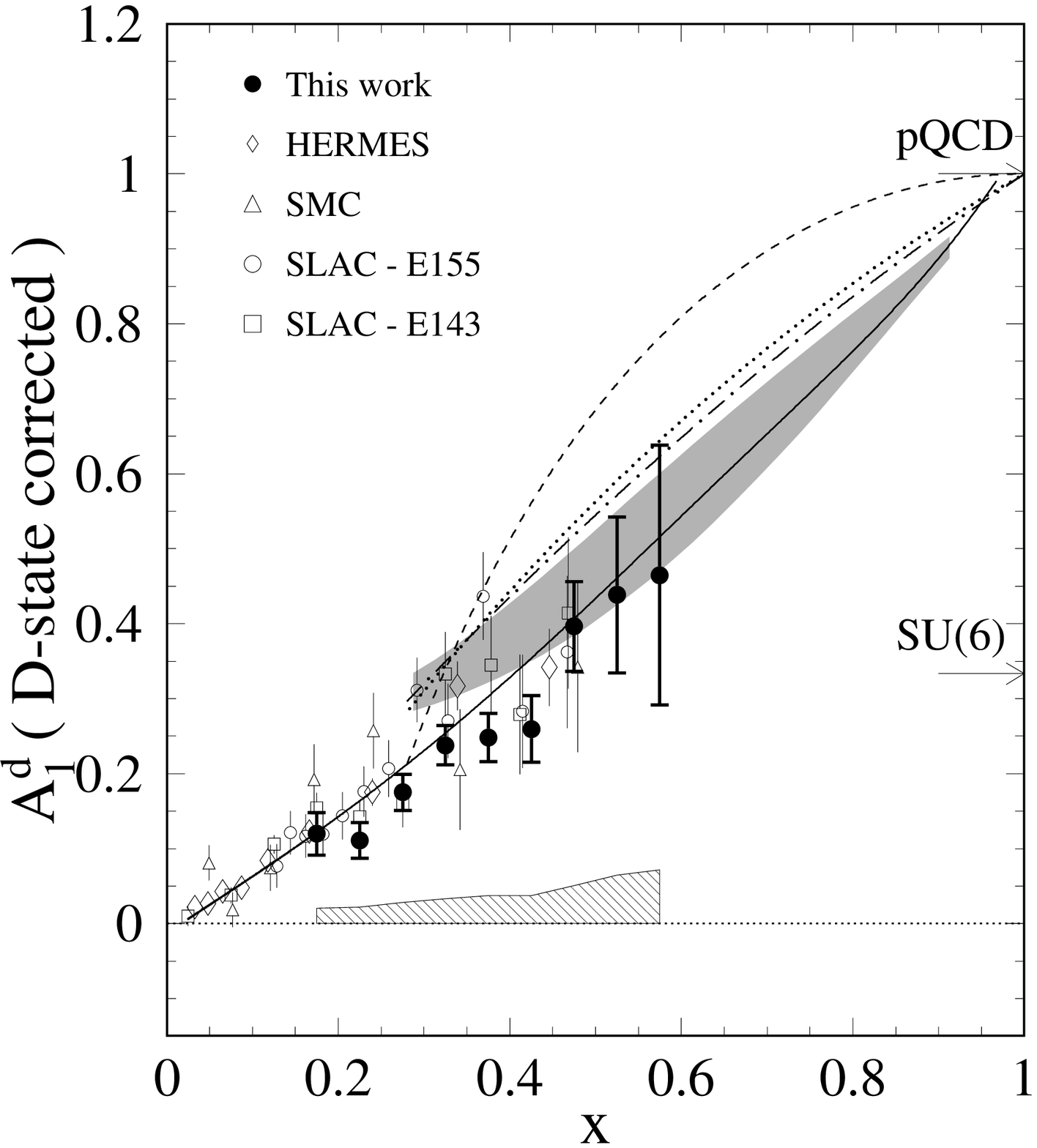}
\end{center}
\addtocounter{figure}{-2}
\caption{\label{fig:CLAS}Asymmetry $A_1$ for the proton (top) and the deuteron (bottom) from CLAS
EG1 for $Q^2>1~{\rm GeV}^2$ and $W>2~{\rm GeV}$ \cite{Griffioen}.
}
\addtocounter{figure}{1}
\end{minipage}
\end{figure}

Apart from the contribution of the quark spins $\Delta\Sigma$, the nucleon spin sum rule
\[ 
\frac{1}{2} = \frac{1}{2} \Delta \Sigma  + \Delta G + L_z
\]
receives contributions from gluon spins $\Delta G = \int {\rm d}x \, \left( G_+ - G_- \right)$
and from orbital angular momentum $L_z$, which must compensate for the smallness of $\Delta\Sigma$.
Experiments start to obtain information on the gluon polarization $\Delta G$, although 
uncertainties are still large. The gluon polarization can be studied in polarized 
DIS and SIDIS and in $\vec{p}\vec{p}$ interactions.

Inclusive DIS is sensitive to $\Delta G(x)$ through the $Q^2$ 
evolution of $g_1$. 
However, the lack of a polarized lepton--proton collider limits the kinematic range of $g_1$ to 
the fixed-target domain at moderate $x$ and $Q^2$.
The status of QCD fits to the world $g_1$ data from CERN, DESY, JLAB and SLAC 
were reviewed by J.~Bl\"umlein. He also summarized the status of $\alpha_s(M_Z^2)$ as
obtained from DIS up to NNNLO for the unpolarized case and NLO for the polarized case 
\cite{Blumlein:2006be}. 
Although the precision of $\alpha_s$ from polarized DIS can not yet reach that from the unpolarized data,
the precision is remarkable.
 
\begin{figure}[tb]
\includegraphics[width=0.48\columnwidth]{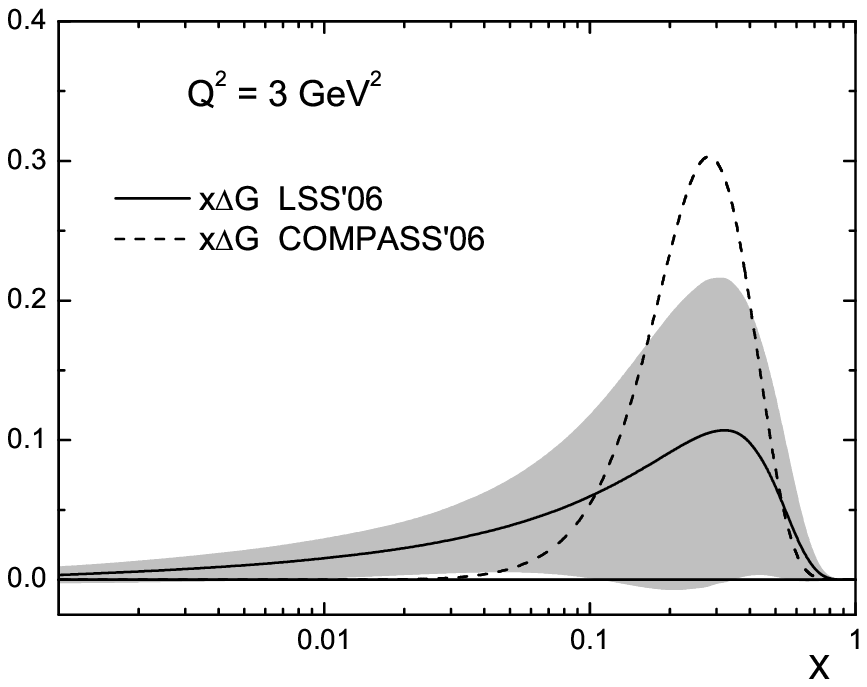}\hfill
\includegraphics[width=0.48\columnwidth]{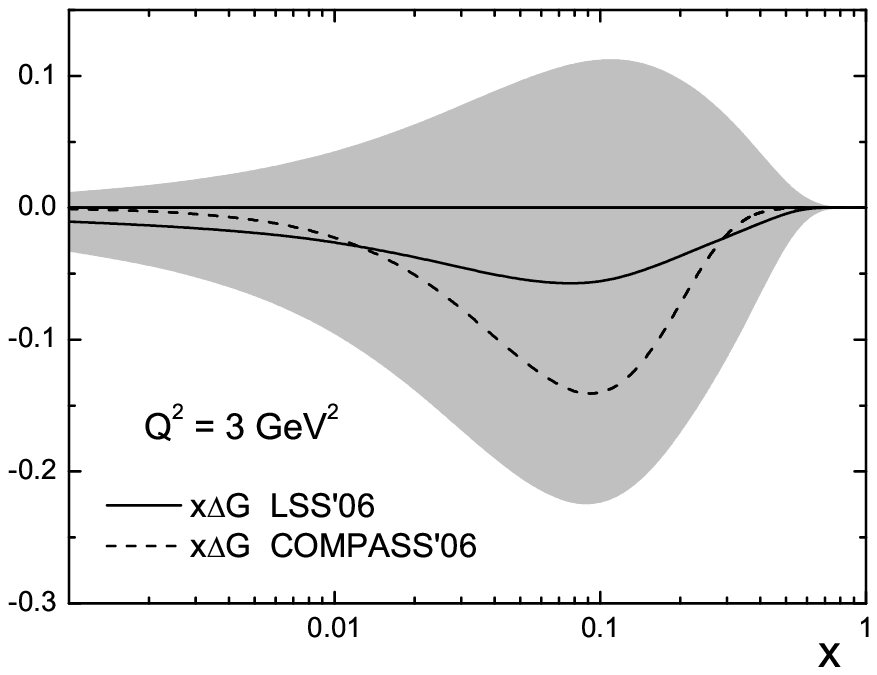}
\caption{\label{fig:LSS}Solutions for positive (left) and negative (right) gluon polarizations 
$x\Delta G(x)$ as function of $x$ from LSS \cite{Leader:2006xc} (solid line) with uncertainties. 
The dashed line shows the corresponding COMPASS fits \cite{Alexakhin:2006vx}.}
\end{figure}

As example for the present status of the QCD analyses we show the recent one by the LSS group 
\cite{Leader:2006xc}, which takes into account the latest data from COMPASS \cite{Alexakhin:2006vx} 
and CLAS \cite{Griffioen}.
They obtain three equally good solutions for positive, negative and sign-changing gluon polarization. 
The positive and negative solutions are compared in Fig.~\ref{fig:LSS} with the solutions obtained
by COMPASS in a similar analysis.
At present even the sign of the gluon polarization cannot be determined from DIS data, however all 
fits yield a small value for the first moment $|\Delta G|\simordertwo 0.3$ at $Q^2=3~{\rm GeV}^2$.

At small $Q^2$ standard DGLAP fits cannot be applied without considering higher-twist effects. 
The LSS group explicitly included such terms in their fits. The resulting higher-twist contributions 
are driven by the CLAS data \cite{Griffioen}. Ermolaev focused on
the small-$x$ aspects of the singlet part of $g_1$, in particular the 
resummation of the leading $\ln 1/x$ terms \cite{Ermolaev:2006ba}. He
suggested to study the dependence on the invariant energy $2P\cdot q$ rather
than $Q^2$ to estimate the impact of the initial gluon density.   

More direct information on the gluon polarization can be obtained in SIDIS. 
Photon--gluon fusion (PGF) $\gamma g \rightarrow q\overline{q}$ leading to a quark--antiquark pair 
gives rise to a double-spin cross-section asymmetry proportional to the gluon polarization
$$
A_{\parallel} = R_{\sf pgf}
a^{\sf pgf}_{LL}\frac{\Delta G}{G}+A_{\sf bgd},
$$
where $R_{\sf pgf}$ is the fraction of PGF events and $a^{\sf pgf}_{LL}$ is 
the analyzing power of the PGF subprocess.
For a particular measurement both, $R_{\sf pgf}$ and the average $a^{\sf pgf}_{LL}$, have to be estimated 
using Monte Carlo (MC) simulations. This introduces a model dependence in the determination of $\Delta G/G$.
In the light-quark case the QCD-Compton process $\gamma q \rightarrow qg$ and the direct 
process $\gamma q \rightarrow q$ limit $R_{\sf pgf}$ to about 30\,\%, while for charmed quark pairs $R_{\sf pgf}$
is essentially unity.
Here the challenges are the low production cross-section and the detection of open charm ($D$ mesons).
The most promising decay channel $D\rightarrow K\pi$ has a branching ratio of only 3.8\,\% which implies
that only one of the two charmed hadrons can be observed with reasonable statistics. Until now all analyses were performed in leading order.

\begin{wrapfigure}{r}{0.5\columnwidth}
\centerline{\includegraphics[width=0.45\columnwidth]{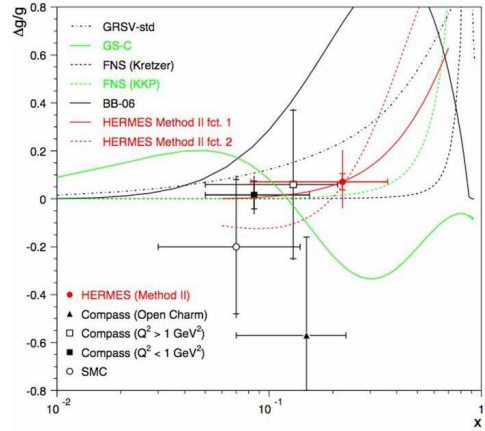}}
\caption{\label{fig:Liebing_Koblitz}The gluon polarization $\Delta G/G$ as 
function of $x$. The new HERMES point lies at $\Delta G/G\simeq0.07$ and 
$x\simeq0.2$, where the two fitted parameterizations intersect.}
\end{wrapfigure}

HERMES determined $\Delta G/G$ from single high-$p_T^{}$ hadron production 
asymmetries in four bins of transverse hadron momentum 
$p_T^{}$ in the range $1.05~{\rm GeV}<p_T^{}<2.5~{\rm GeV}$ using two methods \cite{Liebing}. 
Method~I directly used the above equation for $A_\parallel$ with $R_{\sf pgf}(p_T^{})$ and 
$a^{\sf pgf}_{LL}(p_T^{})$ determined using a PYTHIA Monte Carlo simulation.  
In Method~II two different parameterizations of $\Delta G/G$ were fitted to the measured asymmetries in 
the four $p_T^{}$ bins. The gluon polarization is small and probed around $x\simeq 0.22$ at $\mu^2=1.35~{\rm GeV}^2$. 
The resulting value $\Delta G/G=0.071\pm0.034\,\mbox{(stat.)}\pm0.010\,\mbox{(syst.)}^{+0.127}_{-0.105}\,
\mbox{(model)}$ is shown together with the fitted parameterizations and other data in 
Fig.~\ref{fig:Liebing_Koblitz}.

COMPASS determined $\Delta G/G$ from the cross-section asymmetries for $D$ meson production in
\cite{Koblitz}. This method relies much less on Monte Carlo simulations but is limited in statistical
precision. A neural network was used to estimate $a^{\sf pgf}_{LL}$ from the event kinematics 
on an event-by-event basis. The result $\Delta G/G=-0.57\pm0.41\pm0.17\, \mbox{(syst.)}$ is compatible 
with zero and probes the gluon distribution around $\mu^2=13~{\rm GeV}^2$ and $x=0.15$. 
This is also compatible with the most precise COMPASS result from light-quark pairs at 
$Q^2<1~{\rm GeV}^2$ of $\Delta G/G=0.016\pm0.058\pm0.055\, \mbox{(syst.)}$. 
All results from PGF in DIS are summarized in Fig.~\ref{fig:Liebing_Koblitz} and in Ref.~\cite{gkm_spin2006}.

\begin{figure}[tb]
\includegraphics[width=0.48\columnwidth]{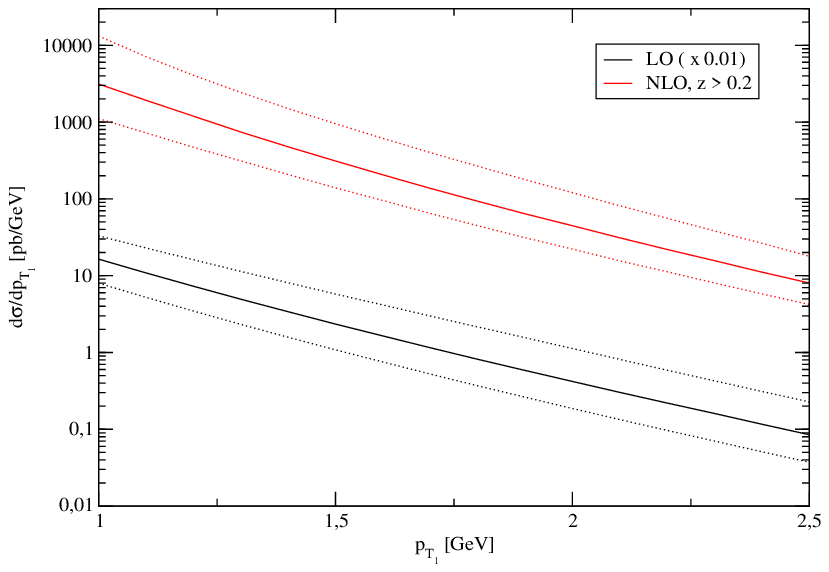}
\hfill
\includegraphics[width=0.48\columnwidth]{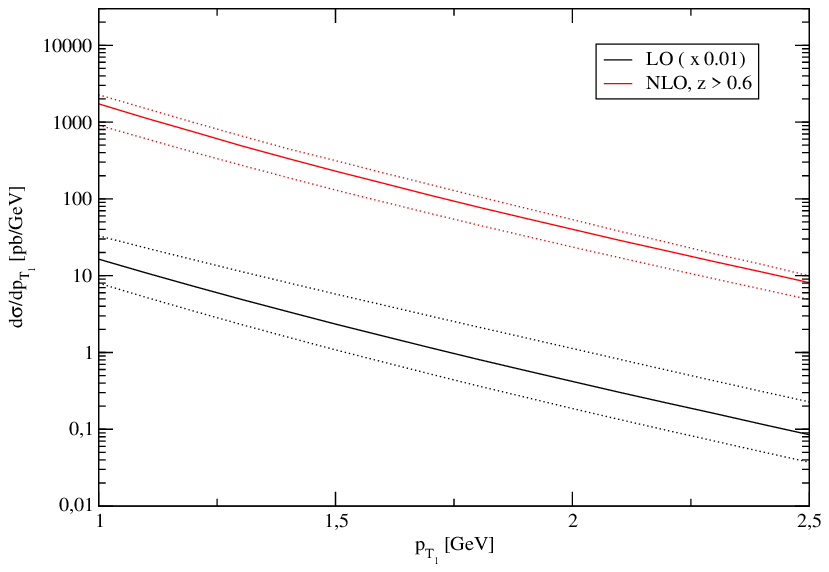}
\caption{\label{fig:Hendlmeier}Scale dependence for the pair cross-section as function of the $p_T^{}$ of the 
first hadron in LO (bottom curves) and NLO (top curves) for $z>0.2$ (left) and $z>0.6$ (right) and COMPASS kinematics.
The scale $\mu$ is varied by a factor two around
$\mu=p_{T,3}^{}+p_{T,4}^{}$, see the text for the definition of $z$.
}
\end{figure}

$\Delta G$ can in principle also be obtained from polarized photoproduction 
of hadron pairs with high transverse momenta ($p_{T,3}^{}, p_{T,4}^{}$). 
Hendlmeier presented NLO calculations for this process with HERMES and 
COMPASS kinematics. The scale dependence for the cross-sections and 
asymmetries at NLO is generally not smaller than at LO. An interesting 
option is the reduction of the scale dependence by cutting on a variable 
defined as $z=-\vec{p}_{T,3} \cdot \vec{p}_{T,4}/\vec{p}_{T,3}^{\,2}$.
This only works for the COMPASS kinematics (Fig.~\ref{fig:Hendlmeier}), 
while in the HERMES case the cut has little effect.

\begin{figure}[tb]
\includegraphics[width=0.54\columnwidth]{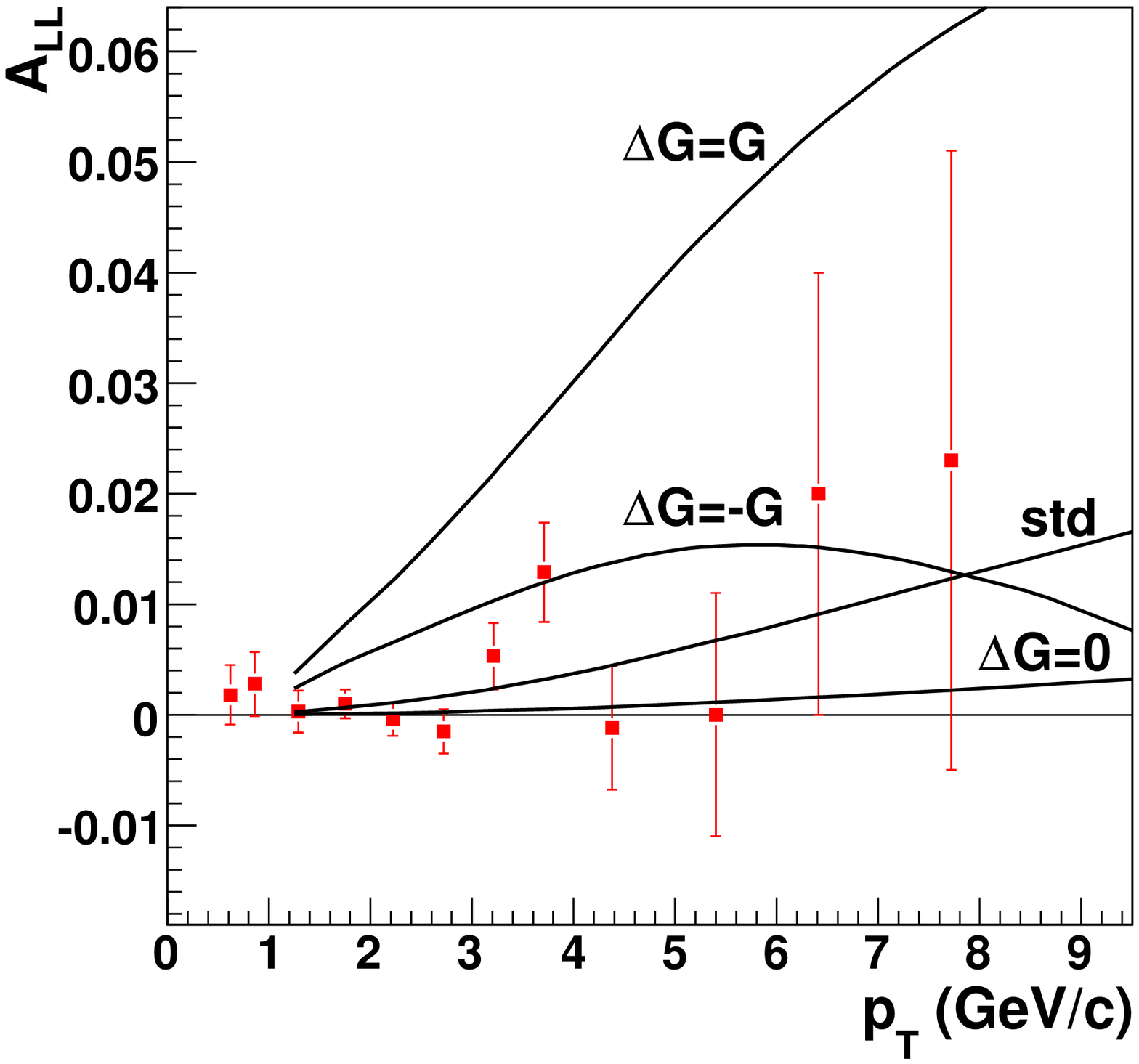}
\hfill
\includegraphics[width=0.44\columnwidth]{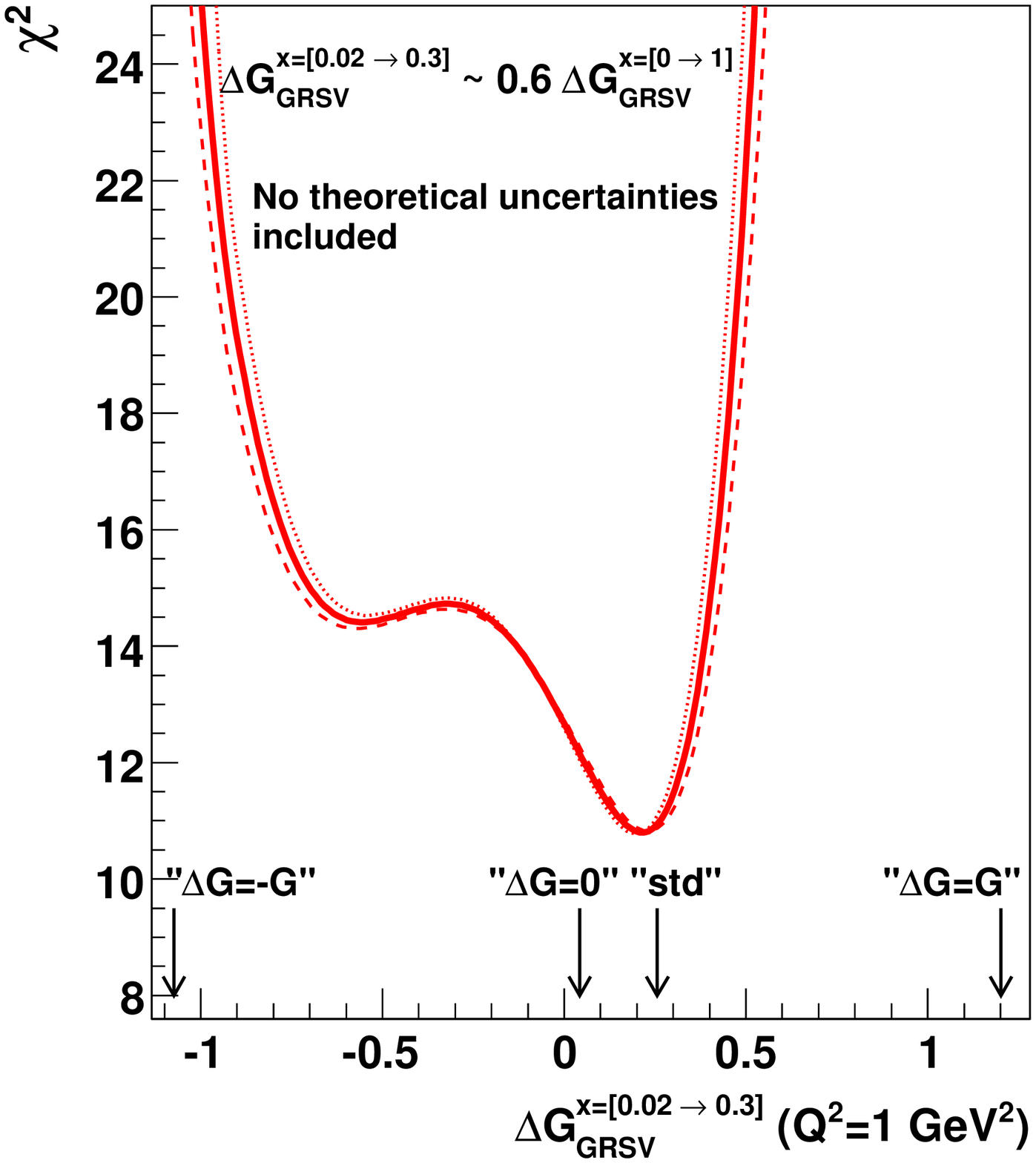}
\caption{\label{fig:Phenix}PHENIX $\pi^0$ asymmetry data \cite{Adare:2007dg}. Left: $A_{LL}$ as function of $p_T^{}$. 
A scale uncertainty of 9.4\,\% is not included. 
The curves correspond to the NLO predictions for various GRSV parameterizations \cite{Gluck:2000dy}.
Right: $\chi^2$ as function of $\Delta G_{\sf GRSV}$, only statistical errors are taken into account.
}
\end{figure}

\begin{figure}[tb]
\begin{minipage}{0.50\hsize}
\begin{center}
\includegraphics[height=0.9\columnwidth,angle=-90]{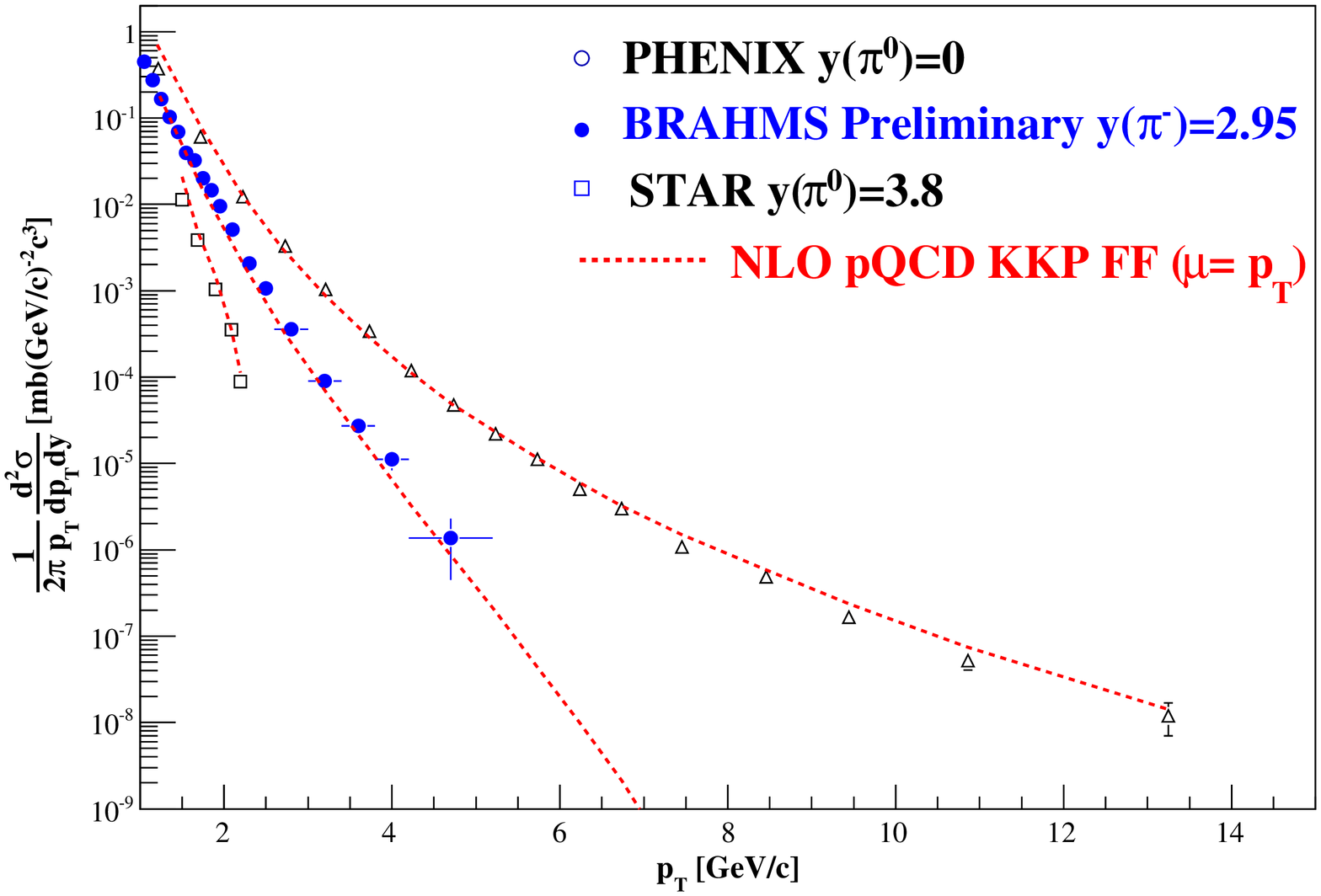}
\end{center}
\end{minipage}
\begin{minipage}{0.50\hsize}
\begin{center}
\includegraphics[width=0.9\columnwidth]{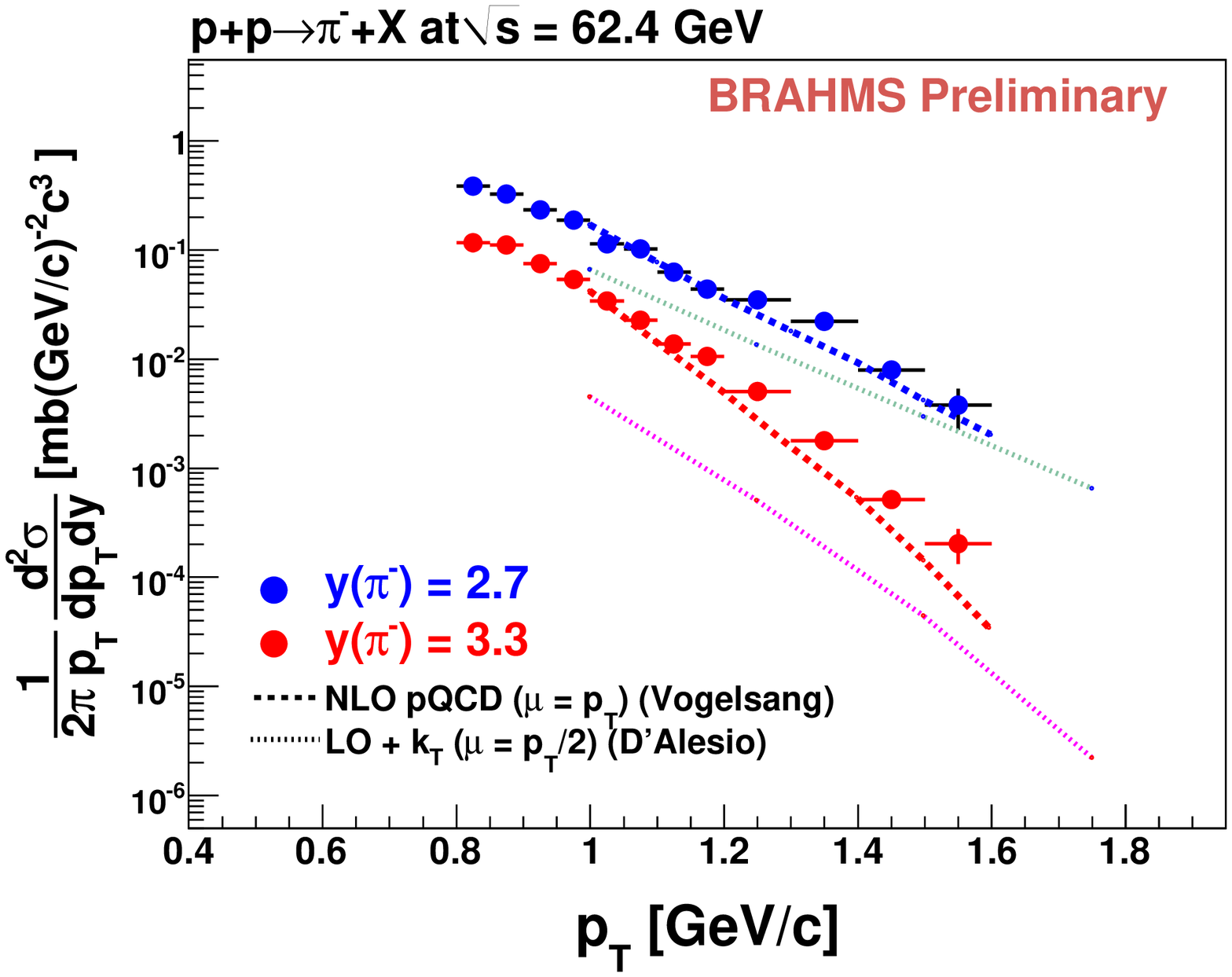}
\end{center}
\end{minipage}
\caption{\label{fig:Lee1} Cross-sections measured at RHIC. Left: Data by the 
PHENIX, STAR and BRAHMS experiments at $\sqrt{s}=200$ GeV, compared with NLO
pQCD predictions. Right: Similarly for data by the BRAHMS experiment at
$\sqrt{s}=62.4$ GeV.}
\end{figure}

At RHIC cross-section asymmetries for longitudinally polarized $\vec{p}\vec{p}$ scattering at 
$\sqrt s=200~{\rm GeV}$ were analyzed for several channels. 
PHENIX presented results for $\pi^0$ production \cite{Adare:2007dg}. 
The cross-section is well understood over seven orders of magnitude in NLO
\cite{Jager:2002xm}, as can be seen in Fig.~\ref{fig:Lee1}. 
Their data favour a small gluon polarization and are compatible with the $\Delta G=0$ and
the standard scenario of GRSV \cite{Gluck:2000dy} in NLO (Fig.~\ref{fig:Phenix}). Also a negative gluon polarization cannot 
be excluded. Future measurements at $\sqrt s = 500$~GeV will remove the present ambiguity because of
the decreasing relative importance of the quadratic term in $\Delta G$ with increasing $p_T^{}$.
Data taken in 2006 at $\sqrt s=62.4$~GeV (see Fig.~\ref{fig:Lee1} for the cross
section measured by BRAHMS at this energy) will allow to probe higher $x_{\sf gluon}$.

\begin{figure}[tb]
\includegraphics[width=0.5\columnwidth]{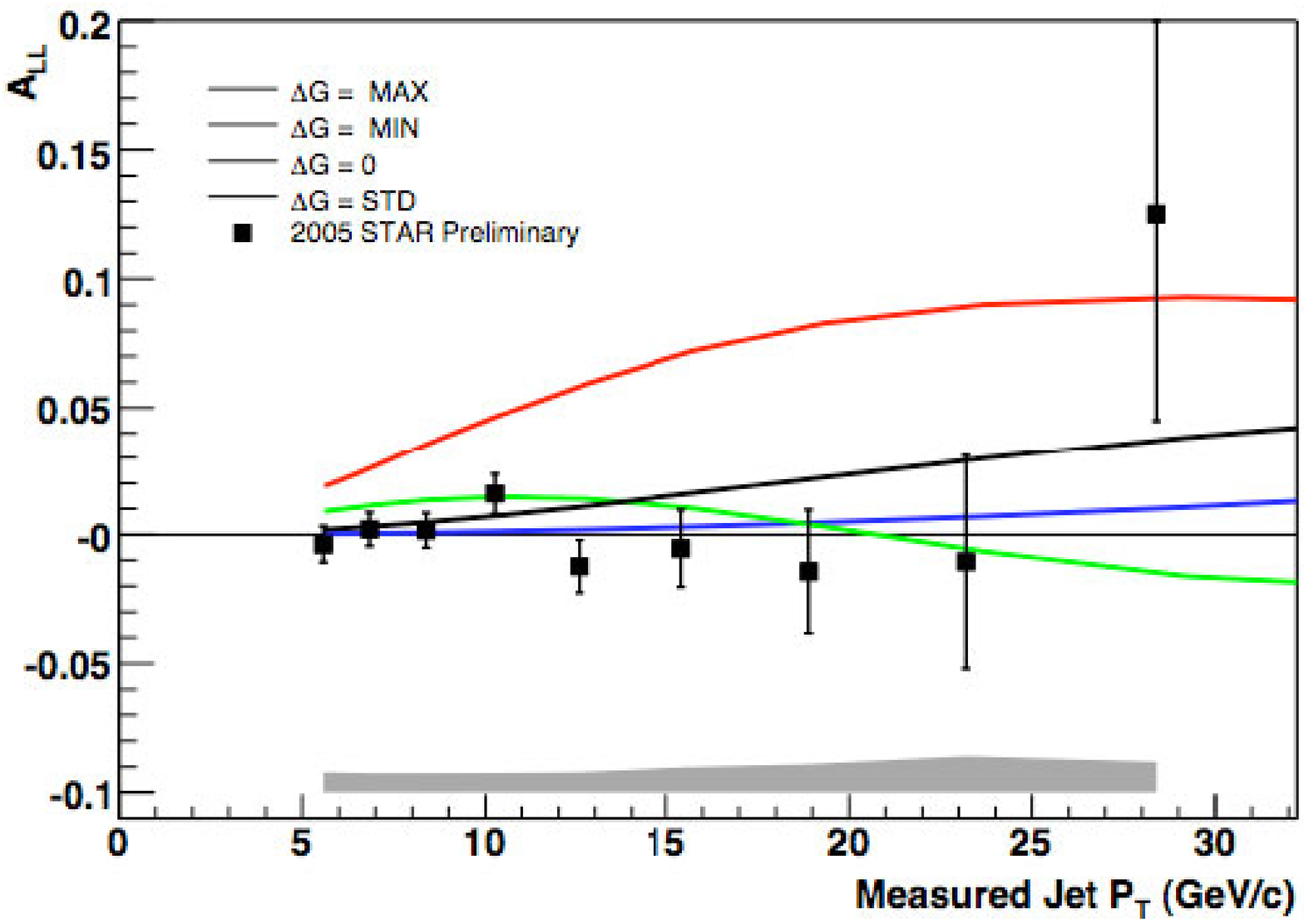}
\hfill
\includegraphics[width=0.44\columnwidth]{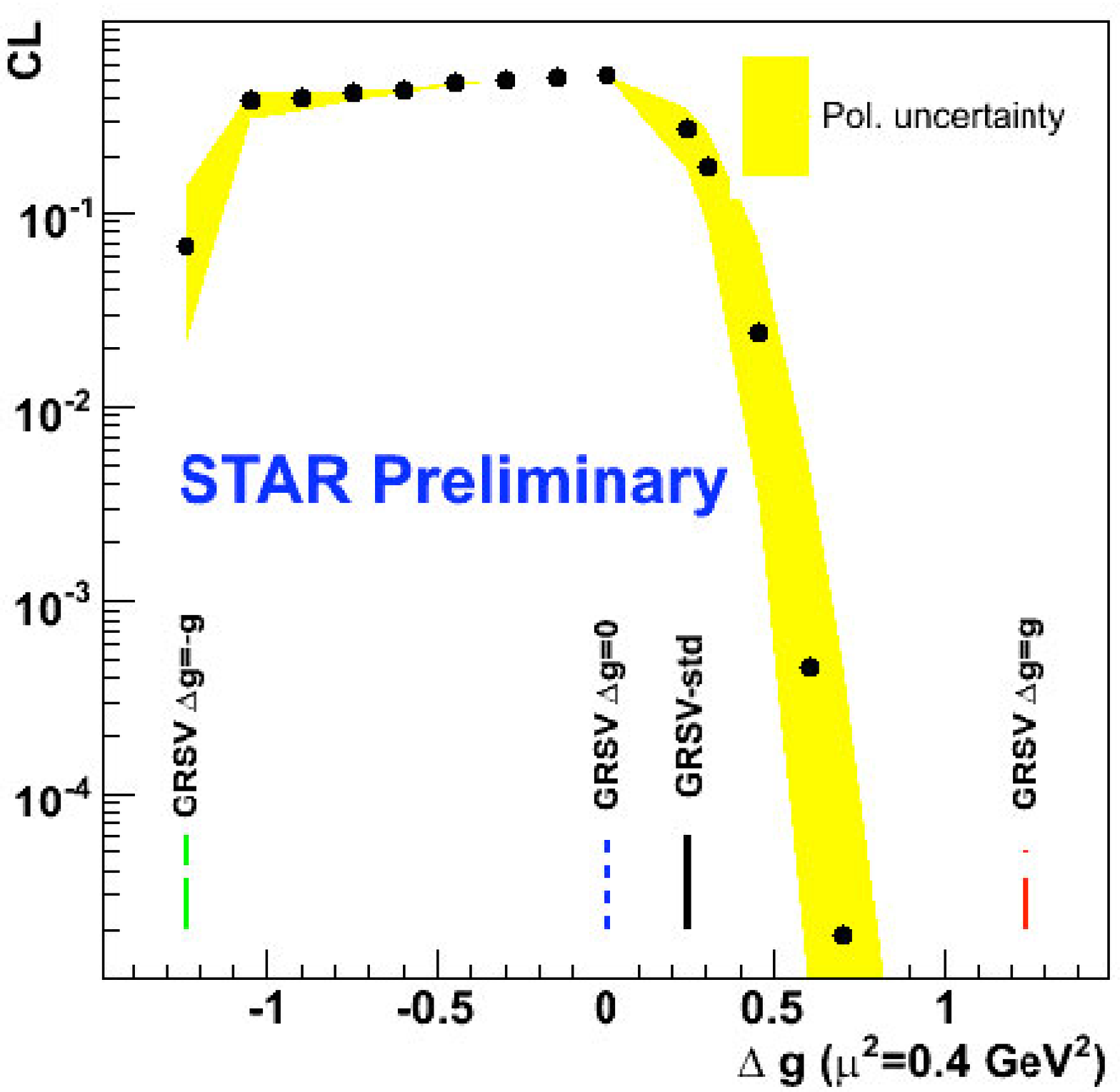}
\caption{\label{fig:Star}STAR inclusive jet asymmetry data. Left: $A_{LL}$ as function of the measured
jet $p_T^{}$. The scale uncertainty of 25\,\% is not included in the shaded systematic error band. 
The curves correspond to the NLO predictions for various GRSV parameterizations \cite{Gluck:2000dy}.
Right: Confidence level as function of $\Delta G_{\sf GRSV}$.
}
\end{figure}

STAR presented longitudinal spin asymmetries for inclusive jet production \cite{Fatemi} (Fig.~\ref{fig:Star}) and 
pions \cite{Simon} from the 2005 run. Again the cross-sections are well understood in NLO \cite{Jager:2004jh} and 
the data point 
to a rather small gluon polarization and negative values cannot be excluded. The precise data taken in
2006 will drastically improve the statistical precision. 

All data suggest gluon polarization $|\Delta G|\simordertwo 0.3$, where one
has to keep in mind that  
the relevant scales for the various measurements vary. 
Although this value is by far smaller than the values around
2--3 predicted by some models assuming a restoration of $\Delta\Sigma$ to the Ellis--Jaffe value of 
0.6 via the axial anomaly, it does not exclude that the gluon and quark spins make up the entire nucleon 
spin of 1/2. 
Therefore, the importance of orbital angular momentum remains to be seen
(further discussion on this topic can be found in section \ref{sec:DVCS}).

\section{Transverse spin}

In analogy to the axial charge $\Delta q$ the tensor charge $\delta q$
is defined as
\[
\langle P,S| \overline{\psi}_q [\gamma^{\mu},\gamma^{\nu}] \gamma_5 
\psi_q(0) | P,S \rangle \sim \delta q \, \left[ P^\mu S^\nu - 
P^\nu S^\mu \right],
\]
which arises for a transversely polarized proton. This fundamental
charge $\delta q$ is the first Mellin moment of the so-called 
transversity distribution $h_1(x)$. It encodes 
completely new information on the proton spin structure and is
difficult, but not impossible to measure. Theoretically the most 
safe extractions can come from processes for which collinear factorization can
be applied. In this case these are the following single and double
transverse-spin asymmetries: 
\begin{itemize}
\item $A_{TT}$ in $p^\uparrow \, \bar{p}^\uparrow \to \ell \bar{\ell} \, X$
\item $A_T$ in various processes exploiting two-hadron fragmentation functions 
\end{itemize}
The HERMES experiment has obtained the first non-zero transversity signal 
from the measurement of  
$A_T$ in the process $e \, p^\uparrow \to e'\, (\pi^+ \pi^-) \, X$
\cite{VanderNat}. 
At DIS 2007 the COMPASS results on 
$A_T$ in $\mu \, d^\uparrow \to \mu'\, (\pi^+ \pi^-) \, X$ were presented: 
they are consistent with zero \cite{Schill}. This is in 
line with the expectation that $h_1^u \approx - h_1^d$ leading to 
cancellations for a deuteron target. In the near future COMPASS will run 
with a proton target, allowing a check of the HERMES results. 
The two-hadron fragmentation functions themselves will be extracted in the
future from BELLE data \cite{Seidl}, which is 
crucial for the quantitative extraction of transversity from 
$e/\mu \, p^\uparrow$ or $p \, p^\uparrow$ processes. On the
theory side, Radici discussed evolution equations for two-hadron fragmentation
functions \cite{Radici}, which is an important issue when extracting 
transversity from a combination of two-hadron production 
observables measured in different experiments. 
Radici pointed out that the $R_T^2$ dependence (which is the square of the
difference of the transverse momenta of the two hadrons) 
leads to a homogeneous evolution equation for the two-hadron
fragmentation functions.  

H\"agler discussed the transverse spin structure of hadrons from
lattice QCD with dynamical quarks, in particular more precise results on
tensor GPDs (generalized parton distributions will be addressed further in 
section \ref{sec:DVCS}) \cite{Haegler}, 
which may also shed light on transverse momentum
dependent parton distributions, as will be discussed below.  

\begin{figure}
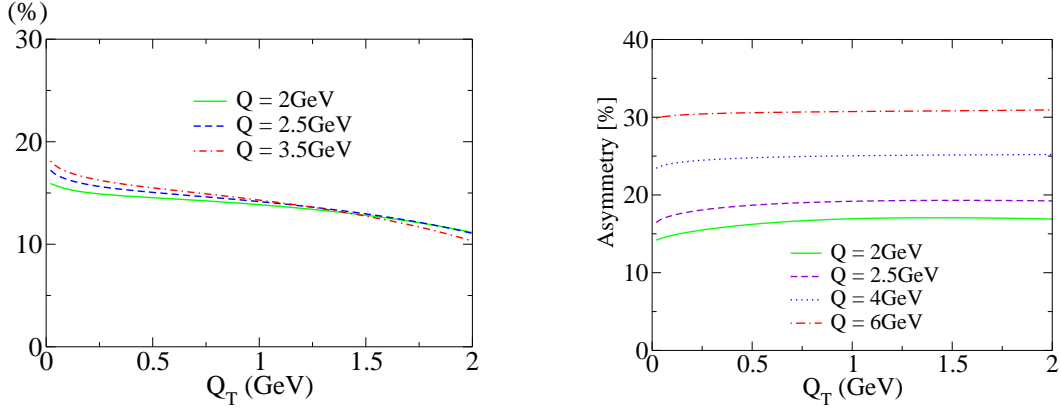

\includegraphics[width=0.45\columnwidth]{plots/Kawamura1.eps}\hfill
\includegraphics[width=0.44\columnwidth]{plots/Kawamura2.eps}
\caption{$A_{TT}(Q_T)$ predictions for $p^\uparrow p^\uparrow$ at
  J-PARC and $p^\uparrow \bar{p}^\uparrow$ at GSI (at $\sqrt{s}=10$ and $14.5$
GeV, resp.) \cite{Kawamura}.}
\label{fig:Kawamura}
\end{figure}

Kawamura presented results \cite{Kawamura} for 
$A_{TT}(Q_T)$ in the Drell--Yan process, which is proportional to $h_1$
times $h_1$. Soft gluon resummation was taken into account. 
Predictions for $p^\uparrow p^\uparrow$ at RHIC and J-PARC and 
for $p^\uparrow \bar{p}^\uparrow$ at GSI were given (Fig.~\ref{fig:Kawamura}). 
The latter observable displays a notably larger dependence
on the scale $Q^2$ than the former. 

Not all transverse spin asymmetries are associated with
transversity though. Large single-spin asymmetries ($A_N$) in 
$p \, p^\uparrow \to \pi \, X$ have been observed by several experiments 
(E704 Collab.~('91); AGS ('99); STAR ('02); BRAHMS ('05); ...). 
The observed asymmetries are left-right asymmetries, which means
the pion distribution is left-right asymmetric depending on the transverse
spin direction and the pion charge. 

\begin{figure}[tb]
\includegraphics[width=0.48\columnwidth]{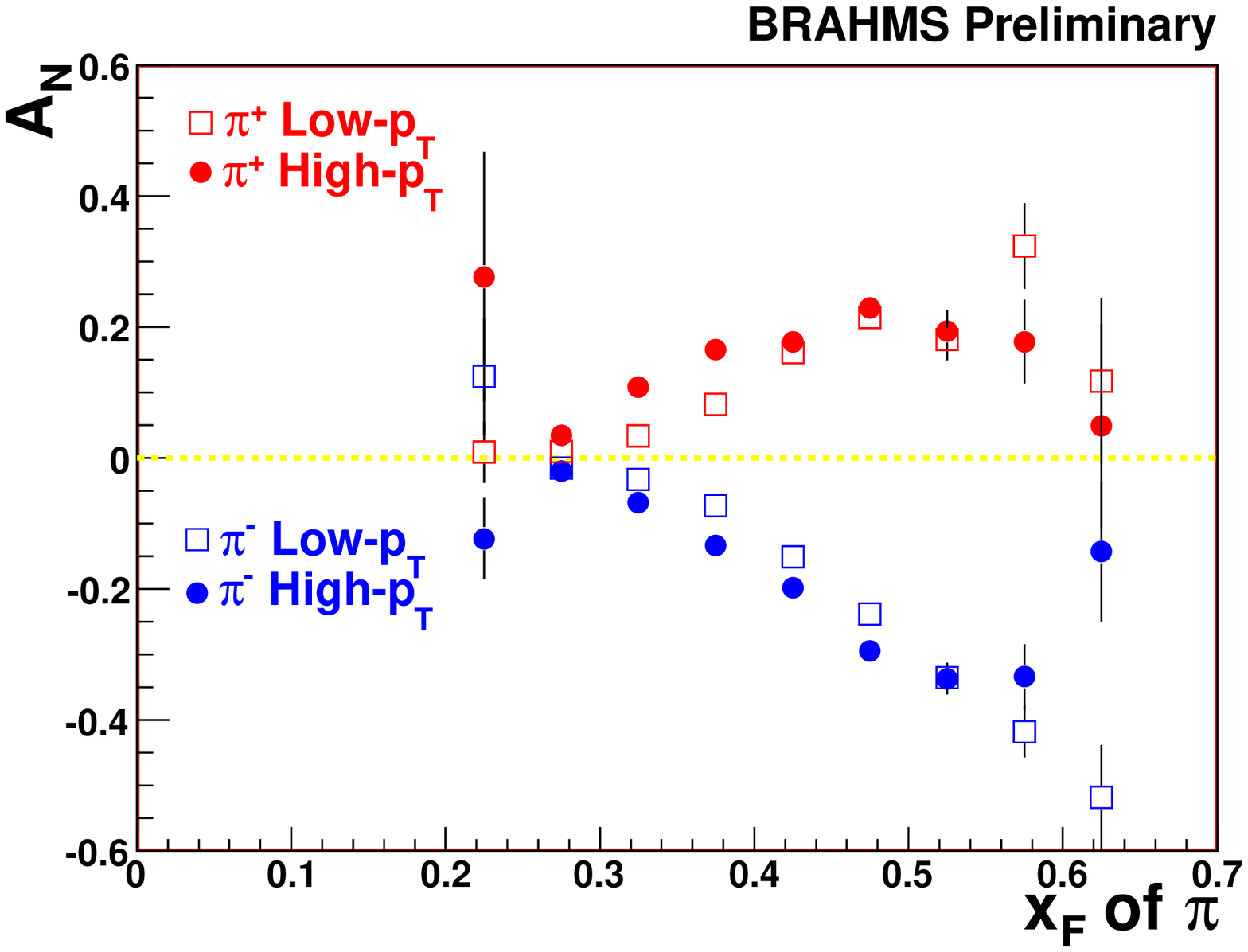}
\hspace{-5 mm}
\includegraphics[width=0.55\columnwidth]{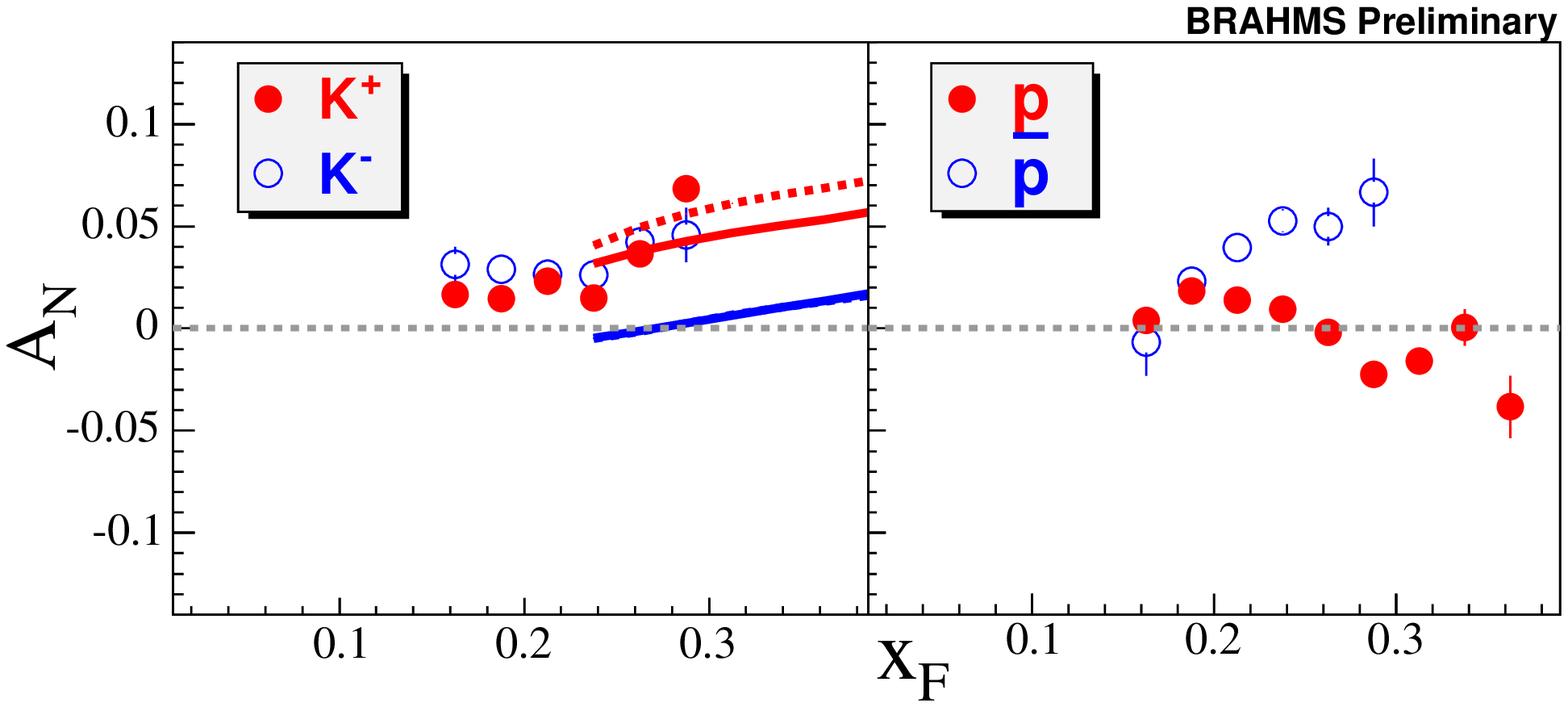}
\caption{\label{fig:Lee2} Single spin asymmetries measured by BRAHMS. Left: 
$A_N$ for $\pi^\pm$ as function of $x_F$, a high
and low $p_T^{}$ data comparison at $\sqrt{s}=62$ GeV. 
Right: $A_N$ for $K^\pm$, $p$ and $\bar{p}$ at $\sqrt{s}=200$ GeV.}
\end{figure}

\begin{figure}[tb]
\includegraphics[width=0.43\columnwidth]{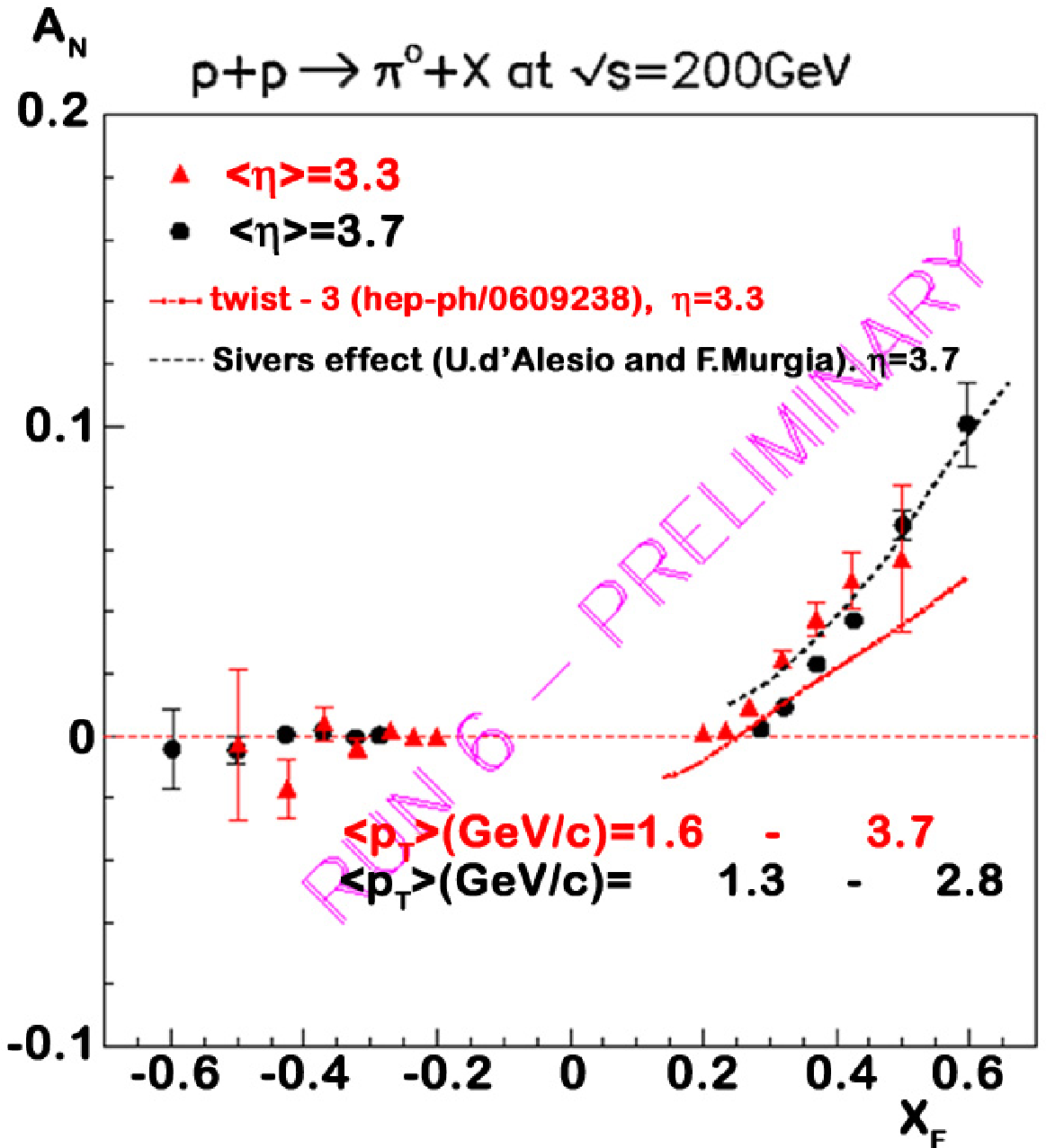}
\hfill
\includegraphics[width=0.53\columnwidth]{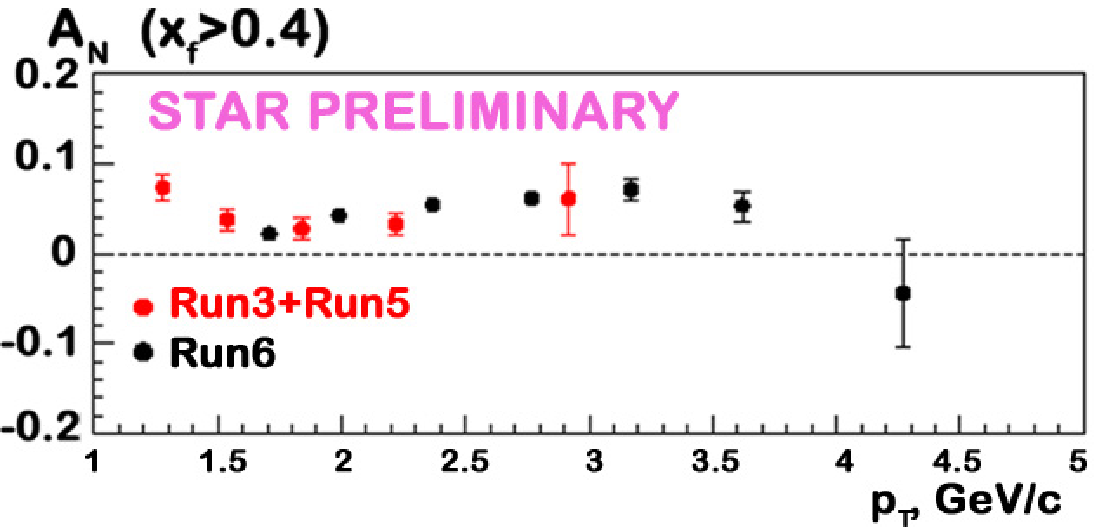}
\caption{\label{fig:Heppelmann} Single spin asymmetries measured by STAR. 
Left: $A_N$ for forward $\pi^0$ as function of $x_F$ at $\sqrt{s}=200$ GeV. 
Right: $A_N$ for $x_F >0.4$ as a function
of $p_T$.}
\end{figure}

New $A_N$ measurements were presented at DIS2007. 
For example, Fig.~\ref{fig:Lee2} shows several 
single-spin asymmetries measured by BRAHMS \cite{Lee}. PHENIX presented  
$A_N$ asymmetries for charged hadrons at mid rapidity as function of $p_T$ 
and for $J/\psi\to \mu^+ \mu^-$ at $x_F \approx \pm 0.1$; all are 
consistent with zero \cite{Eyser}. 
STAR presented $A_N$ asymmetries for forward $\pi^0$'s
and for larger $x_F$ ($>0.4$) also as a function
of $p_T$, these are shown in Fig.~\ref{fig:Heppelmann} \cite{Heppelmann}. 

To understand the origin of these single-spin asymmetries a different 
explanation at the quark--gluon level is required than simply non-zero
transversity.  

One suggestion put forward is to describe $A_N$ at the twist-3 level, the
so-called Qiu-Sterman effect \cite{QiuSterman}. It involves a
matrix element of the form
\[
G_F \sim \langle P,S_T| \; \overline{\psi} (0) \; 
\int \! d\eta^- \; F^{+\alpha} (\eta^-)\; \gamma^+ \, 
\psi(\xi^-) \;| P,S_T \rangle  
\]
This formalism applies at high transverse momentum of the pion. At DIS
2007 recent progress concerning this formalism was presented. 
Koike discussed the recent demonstration of twist-3 factorization and 
gauge invariance of the $A_N$ expression \cite{Koike}. 
Tanaka presented a novel
master formula for $A_N$ in various processes \cite{Tanaka}. 
He showed that the 
twist-3 single-spin asymmetry can be obtained from the twist-2
unpolarized cross-section. This provides a significant simplification 
of the calculation and an understanding of why always the 
combination $G_F - x \, d G_F/d x$ appears. 

Another suggestion is to describe $A_N$ using transverse momentum
dependent parton distributions (TMDs). TMDs arise from the natural
extension of $x$ dependent functions to $x$ and $k_T$
dependent functions. But allowing for a dependence on $k_T$ also
implies the appearance of new functions, such as the Sivers function
\cite{Sivers} $f_{1T}^\perp$:  
\[
f_1(x) \Longrightarrow
f_1(x ,\bm{k}_T^2) + 
\frac{\bm{P} \! \cdot \! \left(\bm{k}_T \times \bm{S}_T \right)}{M} 
\, f_{1T}^\perp (x,\bm{k}_T^2). 
\]
Upon integration over transverse momentum the $k_T$-odd Sivers 
function $f_{1T}^\perp$ drops out. Similarly, a chiral-odd TMD can
arise that is also $k_T$-odd: $h_1^\perp$. In addition, the
fragmentation function analogues $D_{1T}^\perp$ and $H_1^\perp$
arise. 

The Sivers effect can lead to a non-zero $A_N$ in $p \, p^\uparrow \to
\pi \, X$, but also to azimuthal spin asymmetries in many different
processes, such as in semi-inclusive DIS or back-to-back jets in $p\,
p$ scattering. This allows to test the consistency of the many asymmetries
described within this formalism. 

\begin{figure}
\centerline{\includegraphics[width=0.55\columnwidth]{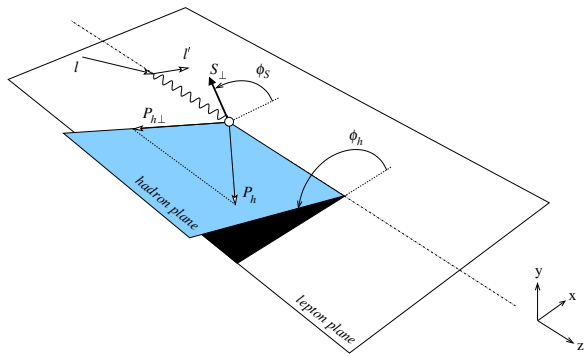}}
\caption{Kinematics of semi-inclusive DIS, with azimuthal angles
  $\phi_S$ and $\phi_h$ indicated \cite{Bacchetta}.}
\label{fig:Bacchetta}
\end{figure}

In semi-inclusive DIS (Fig.~\ref{fig:Bacchetta}) the Sivers function leads to 
a $\sin(\phi_h-\phi_S)$ asymmetry ($\propto f_{1T}^\perp D_1$), which
can be distinguished from the Collins asymmetry $\sin(\phi_h+\phi_S)$
which arises with the transversity function ($\propto h_1 H_1^\perp$) \cite{BoerMulders}. 
Bacchetta presented the complete expressions of all 18 possible semi-inclusive 
DIS structure functions in terms of TMDs \cite{Bacchetta}.

The first azimuthal spin asymmetry measurement was done by the HERMES
Collaboration \cite{HERMES99}. 
At DIS 2007 the latest HERMES and
COMPASS results on the Sivers and Collins asymmetries were presented. 
The HERMES data from 2002--2005 show
large positive (negative) Collins asymmetries for $\pi^+$ ($\pi^-$)
\cite{Diefenthaler},
indicating that the Collins function $H_1^{\perp}$ for favored 
fragmentation is approximately equal in magnitude to unfavored 
fragmentation, but of opposite sign. For the Sivers asymmetry the 
$\pi^+$ data show a significant non-zero
asymmetry, but the $\pi^-$ data are consistent with zero. The neutral pions
follow the expectation from isospin symmetry for both types of
asymmetry. The $K^{\pm}$
asymmetries have less statistical accuracy, but are similar to those
for $\pi^{\pm}$, although $K^+$ shows even larger Sivers 
asymmetries than $\pi^+$. This may
indicate that the sea contribution to the Sivers mechanism is of
importance. COMPASS results on these
and other asymmetries show that for the deuteron these asymmetries 
are all consistent with zero, indicating cancellations rather than
small functions \cite{Bressan,Kotzinian}.

\begin{figure}
\centerline{\includegraphics[width=0.65\columnwidth]{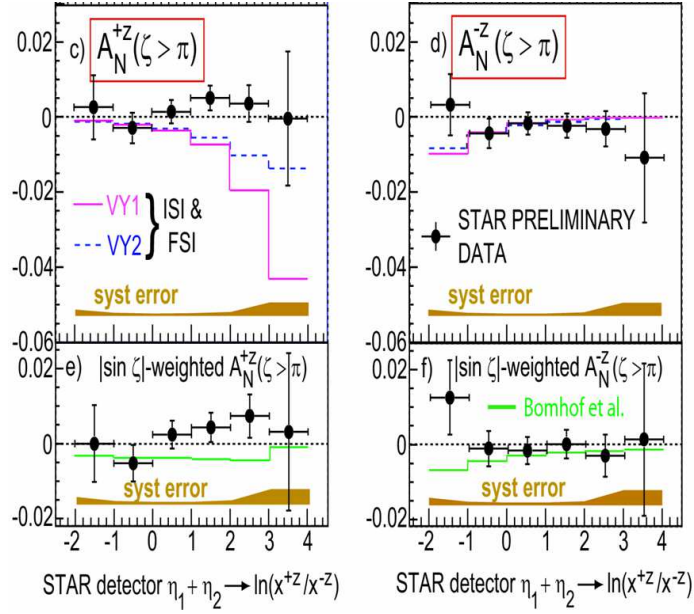}}
\caption{The bisector left-right asymmetry in $p^\uparrow \, p \to \rm{jet} \; 
\rm{jet} \; X$ measured by STAR \cite{Balewski}.}
\label{fig:Balewski}
\end{figure}
As mentioned, the Sivers effect can also lead to a non-zero $A_N$ asymmetry for
back-to-back jet production in $p \, p^\uparrow$ scattering
\cite{BoerVogelsang}. In
general, the two jets are not exactly back-to-back and an asymmetric
distribution of one jet around the other may arise from the Sivers
effect. This effect translates into a (generally smaller) left-right 
asymmetry for the bisector of the two jet directions.   
STAR results on the bisector left-right asymmetry are consistent with zero 
\cite{Balewski}. The data are also consistent with a
recent prediction presented by Bomhof \cite{Bomhof}, 
based on Sivers function input
from semi-inclusive DIS which probes mostly the large-$x$ part of the
Sivers functions. 
One concludes that the smaller $x$ part that is probed in
the back-to-back jets Sivers asymmetry is likely to be small. 
However, another aspect that contributes to the suppression of the 
magnitude of
the back-to-back jets Sivers asymmetry is that the color flow of the
process makes it less sensitive to the Sivers function. It has been
noted several years ago by
Collins \cite{Collins02} that TMDs can exhibit a calculable process
dependence, leading to the result that the Sivers function that enter
the semi-inclusive asymmetry enters the analogous Drell--Yan asymmetry
with opposite sign. This is due to the different color flows in the two
processes. Bomhof and collaborators have found that the more hadrons
are observed in a process, the more complicated the end result. At DIS
2007 Bomhof presented results \cite{Bomhof} for 
$p^\uparrow \, p \to \rm{jet} \;  
\rm{jet} \; X$, included in Fig.~\ref{fig:Balewski}.

\begin{wrapfigure}{r}{0.5\columnwidth}
\centerline{\includegraphics[width=0.45\columnwidth]{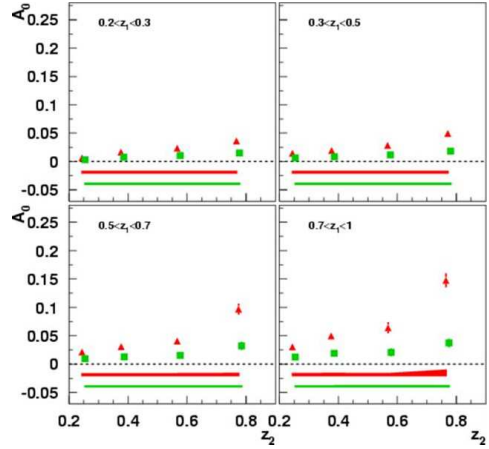}}
\caption{The analyzing power $A_0$ of the $\cos 2\phi$ asymmetry from 547
  fb$^{-1}$ of BELLE data~\cite{Seidl}.}
\label{fig:Seidl}
\end{wrapfigure}

Also the Collins function can lead to asymmetries in other processes
besides semi-inclusive DIS. It leads to $\cos 2\phi$ asymmetries in
several processes, most notably in $e^+ \; e^- \rightarrow \pi^+ \;
\pi^- \; X$, which can be used to extract the Collins function \cite{BJM}. 
This has been done using BELLE data. 
The latest results of this analysis are shown in Fig.~\ref{fig:Seidl} 
\cite{Seidl}, with an impressive factor of 19 more statistics compared to 
the published results \cite{BELLE06}.

\begin{figure}
\centerline{\includegraphics[height=0.65\columnwidth,angle=-90]{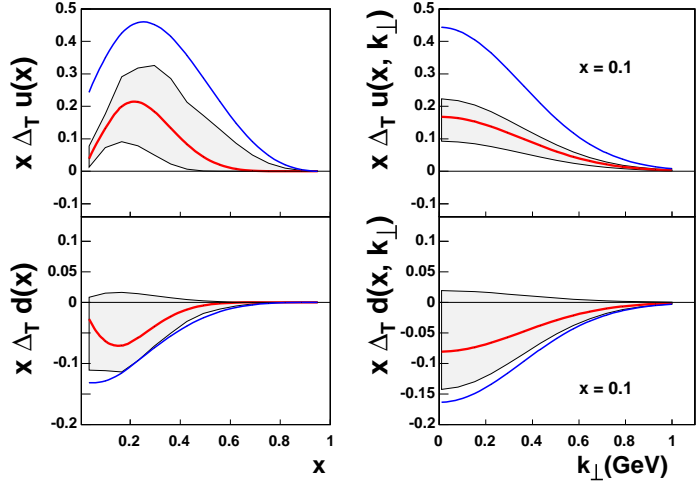}}
\caption{Left panel: the transversity distributions for $u$ and $d$ quarks 
times $x$, as obtained from a transversity and Collins function 
fit to BELLE, COMPASS and HERMES data. Right 
panel: transverse momentum dependence at $x=0.1$. For details 
cf.~\cite{DAlesio}}
\label{fig:DAlesio}
\end{figure}

D'Alesio presented a fit of $h_1$
and the Collins functions
$H_1^\perp$ from both the $e^+e^-$ $\cos 2\phi$ asymmetry (the
published data \cite{BELLE06}) and the semi-inclusive Collins 
$\sin (\phi_h+\phi_S)$ 
asymmetry (using both HERMES and COMPASS data) \cite{DAlesio}. 
It is interesting to see that all this data can be 
simultaneously described within the TMD framework. The result
supports the above-mentioned observation that the Collins function 
for favored fragmentation is approximately equal in magnitude to unfavored 
fragmentation, but of opposite sign. The extracted transversity
functions indicate $|h_1^d(x)| < |h_1^u(x)|$ and opposite sign of
$h_1^u$ w.r.t.\ $h_1^d$, see Fig.~\ref{fig:DAlesio}. The question of how to 
evolve the considered TMD-dependent observables was not yet addressed. 

\begin{wrapfigure}{r}{0.5\columnwidth}
\centerline{\includegraphics[width=0.45\columnwidth]{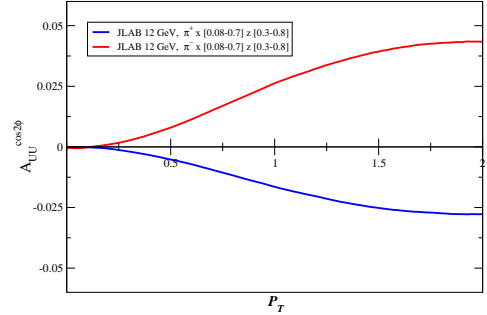}}
\caption{Model prediction of the $\cos 2\phi$ asymmetry in
unpolarized semi-inclusive DIS for JLab@12GeV (updated plot by Gamberg).}
\label{fig:Gamberg}
\end{wrapfigure}

Gamberg presented a model prediction \cite{Gamberg} 
of the $\cos 2\phi$ asymmetry in
unpolarized semi-inclusive DIS ($\propto 
h_1^\perp H_1^\perp$) for the 12 GeV upgrade at JLab, which should provide
access to $h_1^\perp$ (Fig.~\ref{fig:Gamberg}).
 
\begin{figure}
\centerline{\includegraphics[width=0.75\columnwidth]{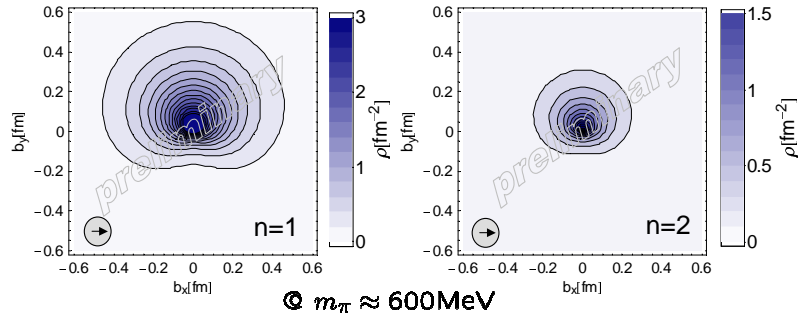}}
\caption{Asymmetric $b_\perp$-space distribution of transversely
  polarized quarks inside a pion from lattice QCD \cite{Haegler}.}
\label{fig:Haegler}
\end{figure}

TMDs like $f_{1T}^\perp$ and $h_1^\perp$ that are odd in $k_T^{}$ are 
spin-orbit coupling quantities, therefore, it is natural to expect a
relation with the orbital angular momentum of the quarks, and hence
with GPDs. Burkardt \cite{Burkardt}
pointed out a model-dependent 
relation between $f_{1T}^{\perp (1)}$ and the GPD $E$ 
\[ 
f_{1T}^{\perp (1)}(x) \propto \epsilon_{i j} S_{T}^{i} b_\perp^j
\int d b_\perp^2 \bm{I}(b_\perp^2)
\frac{\partial}{\partial b_\perp^2} E(x, b_\perp^2)
\]
The factor $\bm{I}(b_\perp^2)$ is not analytically calculable, but 
has to be modeled. Nevertheless, this relation allows to make a qualitative 
link between the Sivers functions and the anomalous magnetic moment of
the $u$ and $d$ quarks. Similarly, Burkardt pointed out a relation
between $h_1^\perp$ and a particular combination of two tensor 
GPDs, for which H\"agler presented
preliminary lattice results from QCDSF/UKQCD \cite{Haegler}, Fig.\ 
\ref{fig:Haegler}. These are the first lattice
results that provide some qualitative information on $h_1^\perp$ of the pion,
indicating that the pion has a surprisingly nontrivial transverse
quark spin structure.
Metz extended this type of model-dependent, but
nontrivial, relations to the other TMDs \cite{Metz}. 

\section{\label{sec:DVCS}Exclusive processes and GPDs}

An outstanding task in solving the 'spin puzzle' of the nucleon is a measurement of
the orbital angular momenta of quarks and gluons.
For the first time, a possibility to reveal the
total angular momentum carried by the quarks in the nucleon~\cite{Ji:1997} 
became available within the formalism of Generalized Parton Distributions (GPDs)
(see~\cite{GPD:reviews} for recent reviews).
These functions are related both to the conventional parton densities and to elastic form factors.
GPDs provide a wealth of new information as they simultaneously measure longitudinal 
momentum distribution and transverse location of partons thereby offering a 
three-dimensional representation of hadrons at the parton level. 

GPDs appear in the scattering amplitude of hard exclusive processes. The DVCS
process, i.e.\ the hard exclusive production of a real photon, provides the 
theoretically cleanest access to GPDs. 
DVCS amplitudes can be measured most readily through the interference between the 
Bethe--Heitler process and the DVCS process.
A large number of reaction channels can be accessed in hard exclusive meson production.
In all cases, polarization observables (e.g.\ single-spin azimuthal asymmetries) are a 
powerful tool to obtain information about GPDs.

{From} the theoretical side, there has been important technical progress in the description of 
hard exclusive processes, with full NLO results in $\alpha_s$ available for most relevant 
channels, partial NNLO results for Compton
scattering and a better understanding of the evolution of GPDs. 
At DIS2007 Diehl presented such NLO calculations
for exclusive meson production at HERA collider and at fixed target kinematics~\cite{Diehl-dis07}.
New avenues have been opened
for the parameterization of GPDs: 
 Luiti introduced an alternative to the mathematical ansatz of
double distributions
in that GPDs are generated from direct constraints from experimental data combined with
lattice calculations yielding a model independent extraction~\cite{Luiti-dis07}.
\begin{figure}
\centerline{
\vtop{\vskip0pt\hbox{%
\includegraphics[bb= 21 13 556 319, clip, width=0.55\textwidth]{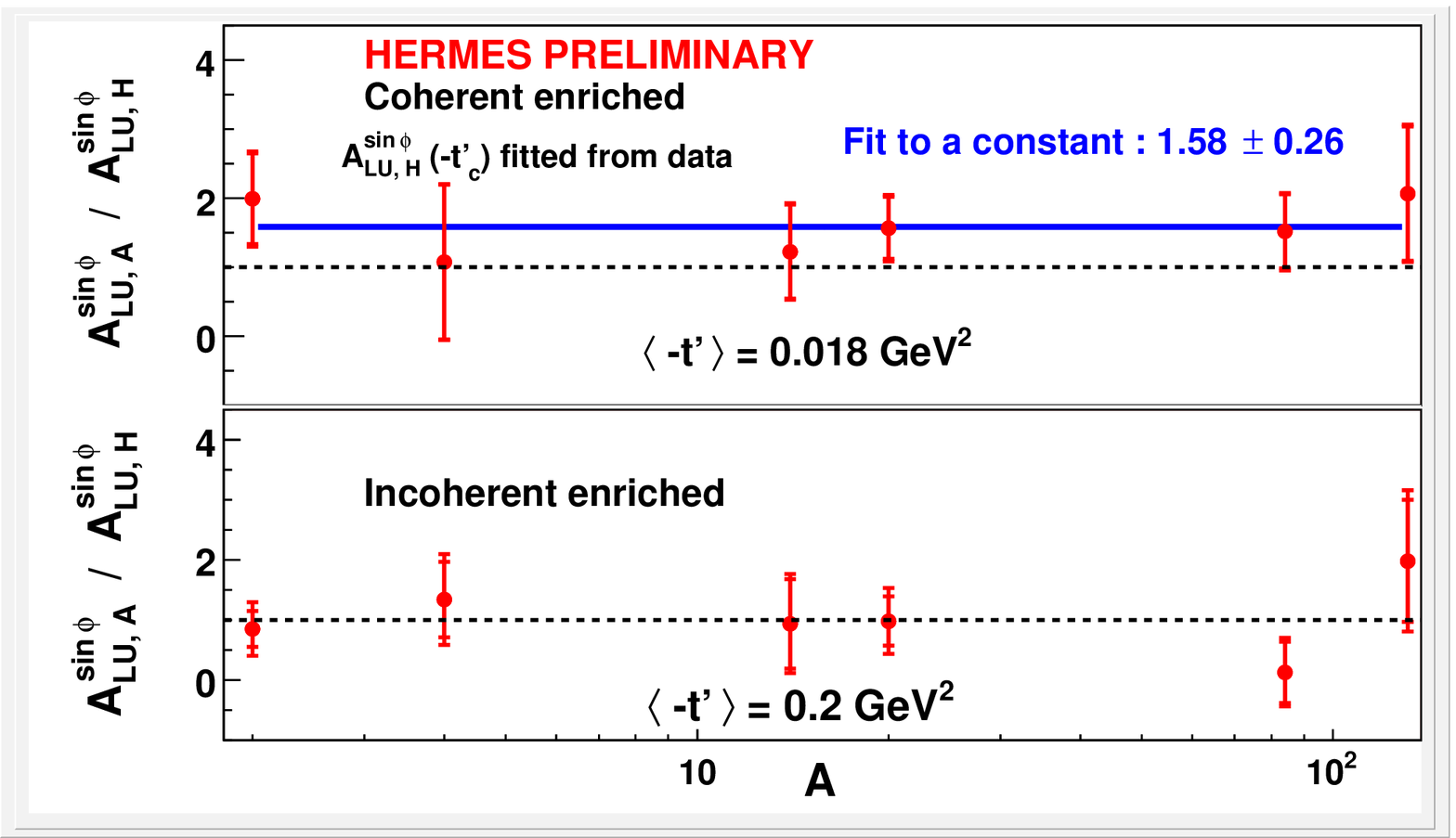}
}}\hspace*{0.2cm}
\vtop{\vskip0pt\hbox{%
\includegraphics[bb= 0 21 566 549, clip, width=0.4\textwidth]{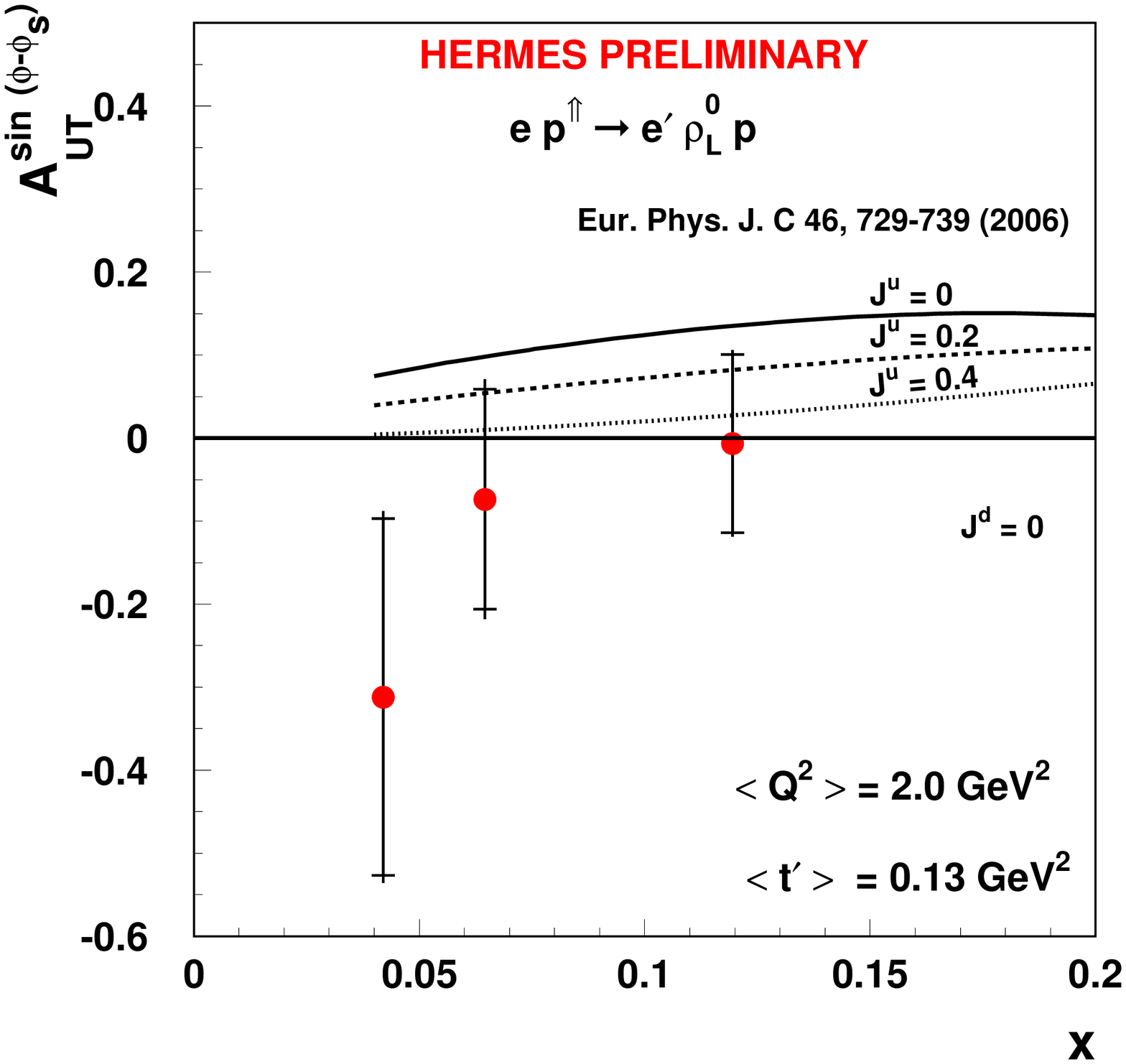}}
}}
\caption{Left panel:
The ratio of the nuclear to free proton DVCS beam-spin asymmetries as a function of the atomic mass
number $A$ measured by HERMES~\cite{Guler-dis07}.
Right panel: Transverse target-spin asymmetry for exclusive production of longitudinally-polarized
$\rho ^0$ measured
by HERMES~\cite{Rostomyan-dis07} and compared to model calculations~\cite{vinnikov}.
}
\label{fig:exclusive}
\end{figure}
Experimental access to GPDs is very difficult as the count rates for 
hard exclusive reactions typically drop drastically with increase of the hardness
of the process.
Nevertheless, there is great progress on the experimental side.
HERMES has presented an overview about the so far measured DVCS observables which
comprises the full set of azimuthal and single-spin asymmetries w.r.t.\ the charge and
helicity of the lepton beam, and w.r.t.\ to the spin polarization of the target, 
either longitudinal or transverse  w.r.t.\ the lepton beam~\cite{Mussgiller-dis07}. 
These results are very promising in view of the greatly improved detection capabilities for
exclusive processes with the information from the recoil detector installed
early 2006.
HERMES also presented the DVCS beam-spin asymmetries measured with a variety of nuclear targets
ranging from  Deuterium to Xenon~\cite{Guler-dis07} which may provide information about
the nuclear forces as well as on the modification of nucleon properties in the nuclear medium.
Fig.~\ref{fig:exclusive}, left panel, shows the ratio 
of the nuclear to free proton DVCS beam-spin asymmetries as a function of the atomic mass
number $A$.
For the coherent region this ratio is
predicted~\cite{Guzey} to have values ranging from 1.85 to 1.95 for $A=12$ to $A=90$.

The Jefferson Laboratory Hall-A experiment presented a measurement of the DVCS cross-section
in the valence quark region on proton and neutron targets~\cite{Voutier-dis07}.
The experiment on the proton provides a strong indication of factorization at
$Q^2$ as low as 2 GeV$^2$, therefore validating a GPD based analysis.

Of particular interest in the context of spin physics   
is the proton helicity-flip distribution $E^q$ which has connection with two crucial 
aspects of spin physics: transverse polarization effects and the orbital angular 
momentum $L^q$  carried by quarks in the nucleon. 
Key observables for these studies are transverse target-spin asymmetries in DVCS and in
exclusive $\rho ^0$ production. HERMES has presented preliminary results for both 
channels~\cite{Rostomyan-dis07,Mussgiller-dis07} (see Fig.~\ref{fig:exclusive} right panel
for the $\rho^0$ asymmetry). Their comparison with a model 
calculation~\cite{VGG} using the total angular momentum of quarks, $J_q$, as input parameter 
in the ansatz for $E^q$ shows that 
these asymmetries are indeed sensitive to $J_u$ in the HERMES kinematics. 
The measurement of the DVCS cross-section on the neutron at Jefferson Laboratory Hall-A  
experiment~\cite{Voutier-dis07} provide information about $J_d$ using the same GPD model.
The complementary constraints on the total angular momenta of up- and down-quarks from both 
experiments remarkably coincide with recent calculations of $J_q$ from lattice QCD~\cite{QCDSF}.   

\section{Conclusion and outlook} 

Exciting new information has been obtained on the nucleon spin structure from polarized 
lepton--nucleon and proton--proton scattering. 
However, a detailed measurement of the gluon polarization remains one of the most important 
issues in spin physics. Running RHIC at higher energy ($\sqrt{s}=500$ GeV) will shed 
more light on this issue.

Transverse spin physics turns out to be a very active and quickly developing field. 
Important results comprise the first extraction of the transversity as well as
of transverse momentum dependent distribution and fragmentation functions like the Sivers 
distribution and the Collins fragmentation function.
These achievements 
can be considered as milestones in the field. They constitute the first step towards a complete
description of the partonic structure of hadrons beyond the collinear parton model. 

A rich future is expected for the elegant concept of generalized parton distributions (GPDs).
Intensive experimental efforts have demonstrated the feasibility of measurements of 
hard-exclusive reactions in a large variety of channels. It turned out that
polarization observables serve as a very powerful tool to access the different GPDs.
The interplay between spin degrees of freedom and parton orbital angular momentum will be a 
key to understand the spin structure of the nucleon.

\vspace{1 cm}

We thank the organizers for the kind invitation to be part of this 
successful workshop and furthermore, all speakers of the spin physics 
sessions for making it such an exciting Working Group.
%

\begin{footnotesize}

\end{footnotesize}


\end{document}